\RequirePackage{etoolbox}
\csdef{input@path}{{style/}{graphics/}}
\documentclass[aos,MSNbibl,nameyear,seceqn,dvips]{arximspdf}

\doi{10.1214/15-AOS1311}% Updated by VTEXPTS2LaTeX.exe, 19.02.2015
\volume{43}
\issue{2}
\pubyear{2015}
\firstpage{769}
\lastpage{784}
\docsubty{FLA}

\makeatletter
\newcommand{\rrvert}{\vert}
\newcommand{\llvert}{\vert}
\newtheorem{teo}{Theorem}[section]
\newtheorem{lemma}{Lemma}[section]
\newproclaim{definition}{Definition}[section]
\newcommand{\conv}{\operatorname{conv}}
\newcommand{\E}{\mathbb{E}}
\newcommand{\D}{\mathcal{D}}
\newcommand{\R}{\mathbb{R}}
\newcommand{\N}{\mathbb{N}}
\newcommand{\M}{\mathcal{M}}
\newcommand{\A}{\mathcal{A}}
\newcommand{\p}{\mathbb{P}}
\newcommand{\X}{\mathcal{X}}
\renewcommand{\phi}{\varphi}
\makeatother

\begin{document}
\begin{frontmatter}

\title{On nonnegative unbiased estimators}
\runtitle{On nonnegative unbiased estimators}

\begin{aug}
% Corresponding author: Pierre Jacob - pierre.jacob@stats.ox.ac.uk% Updated by VTEXPTS2LaTeX.exe, 20.02.2015 09:17
%Updated by VTEXPTS2LaTeX.exe, 19.02.2015 15:05
\author[A]{\fnms{Pierre E.}~\snm{Jacob}\corref{}\thanksref{T1,T2}\ead[label=e1]{pierre.jacob@stats.ox.ac.uk}}
\and
\author[B]{\fnms{Alexandre H.}~\snm{Thiery}\thanksref{T2}\ead[label=e2]{a.h.thiery@nus.edu.sg}}
\runauthor{P. E. Jacob and A. H. Thiery}
\affiliation{University of Oxford and National University of Singapore}
\address[A]{Department of Statistics\\
University of Oxford\\
1, South Parks Road\\
OX1 3TG Oxford\\
United Kingdom\\
\printead{e1}}
\address[B]{Department of Statistics and Applied Probability\\
National University of Singapore\\
Block S16, Level 7, 6 Science Drive 2\\
Singapore 117546\\
\printead{e2}}
\end{aug}
\thankstext{T1}{Supported by EPSRC through Grant EP/K009362/1.}
\thankstext{T2}{Supported by a Singapore Ministry of Education grant.}

%and both authors were supported by a Singapore MoE grant.

% HISTORY:
%
\received{\smonth{5} \syear{2014}}% Updated by VTEXPTS2LaTeX.exe,
%19.02.2015 15:05
%
\revised{\smonth{10} \syear{2014}}% Updated by VTEXPTS2LaTeX.exe,
%19.02.2015 15:05

% ABSTRACT
%
\begin{abstract}
We study the existence of algorithms generating almost surely nonnegative
unbiased estimators. We show that given a nonconstant real-valued function
$f$ and a sequence of unbiased estimators of $\lambda\in\mathbb{R}$,
there is no
algorithm yielding almost surely nonnegative unbiased estimators of
$f(\lambda) \in\mathbb{R}^+$. The study is motivated by
pseudo-marginal Monte
Carlo algorithms that rely on such nonnegative unbiased estimators. These
methods allow ``exact inference'' in intractable models, in the sense that
integrals with respect to a target distribution can be estimated
without any
systematic error, even though the associated probability density function
cannot be evaluated pointwise. We discuss the consequences of our
results on
the applicability of pseudo-marginal algorithms and thus on the
possibility of
exact inference
in intractable models. We illustrate our study with particular
choices of functions $f$ corresponding to known challenges in
statistics, such
as exact simulation of diffusions, inference in large datasets and doubly
intractable distributions.
\end{abstract}

% KEYWORDS
\end{frontmatter}

%s1 #&#
\section{Introduction}
\label{secintroduction}

%s1.1 #&#
\subsection{Exact inference through unbiased estimators}\label{subsecmotivation}

Consider the problem of estimating the integral of a function $\varphi
$ with respect to a
probability distribution with density $\pi$. A successful Markov chain Monte
Carlo or sequential Monte Carlo method allows us to estimate integrals with
respect to $\pi$ in such a way that the error can be reduced down to
zero by producing more samples. We call these
methods ``exact'' since there is no systematic error in the estimation, even
though the sampling error can be large for a given computational budget.
Using the Metropolis--Hastings algorithm, exact inference is possible
when the
target probability density function $\pi$ can be evaluated pointwise
up to a
multiplicative constant.

The possibility of performing exact inference without relying on
evaluations of the target probability density function
is an important open question. A class of exact methods, called
pseudo-marginal Metropolis--Hastings,
has been proposed in \citet{andrieuroberts2009}, generalizing
and validating methods
developed in population genetics [\citet{beaumont2003}] and
lattice quantum chromodynamics [\citet{kennedykuti1985}].
Pseudo-marginal methods rely on nonnegative unbiased estimators of
density evaluations $\pi(x)$ instead of the evaluations themselves.
In a related manner, \citet
{liu1998sequential,delmodoucetjasra2007sequential,fearnhead2008particle,papasp2010,tranissquare2013}
show that
sequential Monte Carlo methods remain exact when the importance weights
are replaced by nonnegative unbiased estimators thereof.

The applicability of exact methods has thus been considerably extended since
estimating $\pi(x)$ is generally easier than evaluating it.
For instance, in the common case where the cost of evaluating the
likelihood function grows at least linearly with
the size of the dataset, pointwise posterior density evaluations become
prohibitive for large datasets but can potentially be estimated using
subsampling [\citet{welling2011bayesian,kleiner2011scalable}].
In state space models, the likelihood involves an intractable integral
over a
latent stochastic process but can be estimated using particle filters
[\citet{andrieudoucetholenstein2010}].
In other \mbox{settings,} the likelihood cannot be evaluated because it
involves an
intractable normalizing constant, such as in ``doubly intractable'' models
commonly found in spatial statistics and graphical models
[\citet{moller2006efficient,Everitt2012,Girolami2013}].
Even for simple models and small
datasets, the use of reference priors for an objective Bayesian
analysis leads
to posterior probability density functions that cannot be evaluated pointwise
[\citet{berger2009formal}] for they involve limits or infinite
sums. In each
case, exact inference can still be achieved through a pseudo-marginal
approach, provided that an appropriate nonnegative unbiased estimator
$\widehat{\pi}(x)$ is
available.

Generic techniques to obtain unbiased estimators from biased ones,
referred to as ``debiasing techniques,'' have been developed
independently in various fields and recently reviewed and generalized
in \citet{McLeish2011}, \citeauthor{RheeGlynn2012} (\citeyear{RheeGlynn2012,RheeGlynn2013}).
The combination of debiasing techniques and pseudo-marginal methods provides a promising
roadmap to perform exact inference in a very general setting.
Unfortunately unbiased estimators $\widehat{\pi}(x)$, as produced by
current debiasing techniques, can take negative
values with positive probability, even if their expectations $\pi(x)$ are
known to be nonnegative.
These negative values prevent the direct use of unbiased estimators
within a pseudo-marginal Markov chain algorithm.
Likewise, standard sequential Monte Carlo methods cannot be directly
implemented when negative values can be encountered.

One might want to avoid the sign problem completely by using unbiased
estimators that only take nonnegative values.
In other words, one might hope to find a debiasing technique which
satisfies a sign constraint.
We propose to study the design of such algorithms. In Section~\ref{subsecdebiasing} we recall the main ideas behind
debiasing techniques and highlight the connection with the Bernoulli
factory [\citet{Keane1994}].
In Section~\ref{subsecapplications} we describe applications in statistics.
In Section~\ref{secexistencegeneral} we present a result stating the
nonexistence of generic schemes to obtain nonnegative unbiased estimators.
In Section~\ref{secexistencecases} we discuss their existence under
additional conditions,
which in practice require additional model-specific information. The
results and further research venues are discussed in Section~\ref{secdiscussion}.

%s1.2 #&#
\subsection{Designing unbiased estimators}\label{subsecdebiasing}

Our results are connected to the literature on debiasing techniques and
Bernoulli factories.
In computational physics, \citet{kuti1982stochastic} uses a
method to unbiasedly estimate some elements of the inverse
of a matrix without fully inverting it, while \citet
{wagner1987unbiased} proposes unbiased estimators of functional integrals;
both methods are inspired by an unpublished scheme of J.~von~Neumann
and S.~M.~Ulam.
A similar idea has been proposed by \citet{rychlik1990unbiased}
for estimating the derivative of a regression function and by
\citet{rychlik1995class}
for kernel density estimation. More recently \citet{McLeish2011}
and then \citeauthor{RheeGlynn2012} (\citeyear{RheeGlynn2012,RheeGlynn2013})
have proposed a general scheme to remove the bias of a sequence of
consistent estimators
$(S_n)_{n \geq0}$ of a quantity $\lambda\in\R$, satisfying
%
%e1.1 #&#
%
\begin{equation}
\label{eqmeanconsistency} \lim_{n\to\infty} \E( S_n ) = \E(S) =
\lambda.
\end{equation}
In equation (\ref{eqmeanconsistency}), the quantity $S$ can either
be thought
of as a random variable that is impossible to generate in finite time
and $S_n$ as an approximation of $S$, or simply as the desired, and
generally unknown value $S=\lambda$. Suppose that one can sample from
$S_n$ for each $n
\geq0$.
Let $N$ be an integer-valued random variable that is independent
of the sequence $(S_n)_{n \geq0}$ and that can take arbitrary large values.
Under mild assumptions, with the convention $S_{-1} =0$, the weights
$w_n = 1 /
\p(N \geq n )$ are such that the random sum
%e1.2 #&#
%
\begin{equation}
\label{eqdebias} Y = \sum_{n=0}^N
w_n \times( S_n - S_{n-1} )
\end{equation}
is an unbiased estimator of $\lambda$. The following result gives
a condition for its second moment to be finite.

%
%th1.1 #&#
%
\begin{teo}[{[Theorem 1 of \citet{RheeGlynn2013}]}] \label{teodebiasing}
Introduce a random variable $S$ with $\E(S) = \lambda\in\R$. Let
$(S_n)_{n\geq0}$ be a sequence of random variables,
let $N$ be an integer valued
random variable that can take arbitrarily large values and set $w_n =
1/\p(N
\geq n)$. Under the condition
%
%e1.3 #&#
%
\begin{equation}
\label{eqvariance} \sum_{n = 1}^\infty
w_n \times\E\bigl( \llvert S - S_{n-1}\rrvert
^2 \bigr) < \infty,
\end{equation}
the random variable $Y = \sum_{n=0}^N w_n \times( S_n - S_{n-1}
)$,
with the convention $S_{-1}=0$, is well defined, has expectation $\E
(Y) = \E(S) = \lambda$ and
a finite second moment
\[
\E\bigl(Y^2\bigr) = \sum_{n=0}^{\infty}
w_n \times\bigl( \E\bigl( \llvert S-S_{n-1}\rrvert
^2 \bigr) - \E\bigl( \llvert S-S_{n}\rrvert
^2 \bigr) \bigr) < \infty.
\]
\end{teo}

The ``debiased'' estimator $Y$ also generalizes the random truncation approach
discussed in \citet{papasp2011bayesian,Girolami2013} and
references therein.
The random variable $N$ could be replaced by a stopping time.
Since the random sum in equation (\ref{eqdebias}) only involves an almost
surely finite number of terms, the estimator $Y$ is straightforward to
simulate.

In the case where the quantity of interest $\lambda$ is nonnegative, the
random sum in equation (\ref{eqdebias}) can still take negative
values, even
if the original estimators $(S_n)_{n \geq0}$ were all almost surely
nonnegative; this is because each increment $ (S_n - S_{n-1}
)$ can potentially be negative.
An important exception occurs when the estimators $(S_n)_{n \geq0}$ are
ordered, that is, $S_n \geq S_{n-1}$ almost surely.
With exact inference in mind, one can wonder about the existence of
other debiasing techniques which, unlike
$Y$ of equation (\ref{eqdebias}), would only yield nonnegative values.
Section~\ref{secexistencegeneral} will introduce a framework to
study that question.

Our framework will also be related to Bernoulli factories,
which have been introduced in the seminal article of \citet{Keane1994}.
Given a subset $\mathcal{P}\subset[0,1]$ and a function
$f\dvtx\mathcal{P}\to[0,1]$, a Bernoulli factory
generates Bernoulli random variables with success probability $f(p)$
given as input an independent sequence of Bernoulli random variables
with success
probability $p \in\mathcal{P}$; of course the algorithm does not have access
to the value $p$.
The existence of such an algorithm depends on the subset
$\mathcal{P}$ and on the function $f$ considered.
For instance, there does not exist an algorithm for $f\dvtx p\mapsto2p$ and
$\mathcal{P}=[0,1/2]$; maybe surprisingly, there does exist an
algorithm for
the same function $f$ and the set $\mathcal{P}=[0,1/2 - \varepsilon]$
for any
$\varepsilon> 0$.
It will become apparent in Section~\ref{subseccaseinterval} that the
construction of nonnegative unbiased estimators
shares many similarities with the Bernoulli factory.

To summarize, debiasing techniques allow us to construct unbiased
estimators of generic quantities
but do not ensure that the resulting estimates are nonnegative. On the
other hand Bernoulli factories always produce nonnegative variables,
but require
Bernoulli variables as input. In general we are interested in the
existence of algorithms producing nonnegative unbiased estimators
of $f(\E[X])$ for some function $f\dvtx\R\to\R^+$ and real-valued
random variables $X$.

%s1.3 #&#
\subsection{Applications}\label{subsecapplications}

Some specific choices of function $f$ are of special interest in
applied probability and statistics,
especially the exponential $f\dvtx x\mapsto\exp(x)$ and the inverse
$f\dvtx x\mapsto1/x$.

The exponential case appears whenever log-likelihood evaluations can be
unbiasedly estimated.
An algorithm generating unbiased estimates of $\exp(\lambda)$ from a
stream of unbiased estimates of $\lambda$ is referred to as a \textit
{Poisson estimator} in the literature on perfect simulation and
inference for diffusion processes [\citet
{beskos2005exact,beskos2006exact,beskos2006retrospective,jourdain2007exact,fearnhead2008particle,olsson2011particle,sermaidis2012markov}],
and the first occurrence might be in \citet{bhanot1985bosonic}.
On a finite interval $[0,T]$, the probability distribution
$\mathbb{Q}$ on the space of continuous functions $\mathcal{C}([0,T],
\mathbb{R})$ generated by a scalar diffusion processes with unit volatility
coefficient $dX_t = \mu(X_t) \,dt + dW_t$ has, under mild regularity
assumptions on the drift function $\mu\dvtx\mathbb{R} \to\mathbb{R}$, a
Radon--Nikodym derivative with respect to the standard Wiener measure
$\mathbb{W}$ that can be expressed as
\begin{eqnarray*}
\frac{d \mathbb{Q}}{d \mathbb{W}} \bigl((x_t)_{t=0}^T \bigr)
&=& \exp\biggl( \int_{t=0}^T \Phi(x_t)
\,dt \biggr)
\end{eqnarray*}
for an explicit function $\Phi\dvtx\mathbb{R} \to\mathbb{R}$ given by
Girsanov's
theorem. As described in \citet{beskos2006exact}, unbiased
estimates of the
integral $\int_{t=0}^T \Phi(x_t) \,dt$ can be obtained by standard
importance sampling. The existence of a Poisson estimator allows us to transform
these samples into an unbiased estimate of
$ ( d \mathbb{Q} /d \mathbb{W} ) ((x_t)_{t=0}^T
)$, which can then
be used for exact inference.

The exponential case also appears in the context of inference for
large datasets, where the posterior probability density function $\pi$ is
expensive to evaluate point-wise.
Indeed the log-likelihood
$\ell(\theta) = \sum_{i=1}^n \log f(y_i \mid\theta)$ of $n \gg
1$ independent observations $(y_i)_{i=1}^n$ can be unbiasedly estimated
at reduced
cost by using a random
subsample of only $m \ll n$ observations. For instance, given any $m
\geq1$, the
quantity $\widehat{\ell}(\theta)=(n / m) \sum_{i=1}^m \log
f(y_{\sigma_i} \mid
\theta)$, where $(\sigma_i)_{i = 1}^m$ are drawn uniformly in $\{
1,\ldots,n\}$, is an unbiased estimator of $\ell(\theta)$.

The choice $f\dvtx x\mapsto1 / x$ appears in the context of
doubly intractable models [\citet
{walker2011posterior,Girolami2013}] where the observations are
assumed to follow a distribution with density
\[
f(y\mid\theta) = \frac{g(y,\theta)}{\int g(s,\theta) \,ds}
\]
for a function $(y,\theta) \mapsto g(y,\theta)$ that can be evaluated
pointwise.
The denominator $Z(\theta) = \int g(s,\theta) \,ds$ is generally intractable,
which prevents the use of the standard Metropolis--Hastings algorithm
to obtain
posterior estimates.
Nevertheless $Z(\theta)$ can be unbiasedly estimated by standard
importance sampling. Would a nonnegative estimator of $1 /
Z(\theta)$ be available, a pseudo-marginal approach could be implemented.

%s2 #&#
\section{Existence of nonnegative unbiased estimation schemes}\label{secexistencegeneral}

%s2.1 #&#
\subsection{Algorithms and factories}\label{secexistencedef}

For any nonempty measurable space $\X\subset\R$, let $\M_1(\X)$
be the set of
probability distributions on $\mathcal{X}$ with finite first moment and
$\conv(\X)$ the smallest interval containing $\X$. For $\mu\in
\M_1(\mathcal{X})$ we use the notation $m_1(\mu) = \int_\mathcal
{X} x
\mu(dx)$ for the mean of $\mu$; indeed, $m_1(\mu) \in\conv(\X)$
for any $\mu
\in\M_1(\X)$. The distribution of the random variable $X$ is denoted by
$\D(X)$. Let $L^2(\X)$ be the space of square integrable random
variables on $\X$.
The indicator function of a set $A$ is denoted by $1_A$, and $1_x$ for
some $x\in\R$
denotes the Dirac delta function centered at $x$.
An unbiased estimator of a quantity $\lambda\in\R$ is called a
$U$-estimator of $\lambda$,
or a $U^+$-estimator in the case where it is almost surely nonnegative.

For a function $f\dvtx\conv(\X) \to\mathbb{R}^+$, we propose to study
the existence of \mbox{$f$-}factories,
defined as devices taking as input $U$-estimators of
$\lambda\in\conv(\X)$ with support on $\X$, and producing
$U^+$-estimators of $f(\lambda)$.
Borrowing ideas from \citet{Keane1994}, we first define
rigorously a class of
algorithms that we will consider practical.
%de2.1 #&#

\begin{definition}
\label{defalgorithm}
Let $\X$ be a subset of $\R$. An $\X$-algorithm $\A$ is a pair
$ ( T, \phi
)$ where $T=(T_n)_{n \geq1}$ is a sequence of functions $T_n\dvtx
(0,1) \times
\X^n \to\{0,1\}$, and $\phi= (\phi_n)_{n \geq1}$ is a sequence of functions
$\phi_n\dvtx (0,1) \times\X^n \to\R^+$.
\end{definition}

An $\X$-algorithm $\A\equiv(T,\phi)$
takes an infinite sequence $x=(x_n)_{n \geq1} \in\X^{\infty}$ and
an auxiliary variable $u \in(0,1)$ as input and produces as output
\[
\A(u,x) = \phi_\tau(u,x_1, \ldots,x_\tau)
\]
with $\tau=\tau(u,x)=\inf\{n \geq1\dvtx T_n(u,x_1, \ldots, x_n)=1\}$.
We adopt
the convention $\A(u,x)=\infty$ when $\{n \geq1\dvtx T_n(u,x_1, \ldots,
x_n)=1\}
= \varnothing$ and say in this case that the algorithm does not terminate.
In the applications that we have in mind, the infinite sequence
$x=(x_n)_{n \geq1} \in\X^{\infty}$ is the realization of an
independent sequence
of random variables $X=(X_n)_{n \geq1}$, and the variable $u \in
(0,1)$ is the
realization of a random variable $U \sim\operatorname{Uniform}(0,1)$
independent of $X$.
In this case, we say that~the algorithm almost surely terminates if $\p
(\tau<
\infty)=1$.
Definition \ref{defalgorithm} translates the~fact that a valid algorithm uses a possibly random amount of
inputs and that the decision to stop acquiring more inputs only relies
on the
information contained in the already acquired inputs.

The variable $U$ allows the algorithm to be randomized: on top of the sequence
$(X_n)_{n\geq1}$ it can sample additional random variables.
Specifying a single
auxiliary variable $U \sim\operatorname{Uniform}(0,1)$ or an infinite
independent sequence $(U_n)_{n \geq1}$ of uniforms is equivalent.
Indeed, one can
construct an
infinite sequence of independent Bernoulli random variables by
considering the
binary expansion of $U \sim\operatorname{Uniform}(0,1)$, and then partition
the expansion into
disjoint infinite subsequences to obtain an infinite number of binary
representations of independent uniform random variables.

% DEFINITION
%
%de2.2 #&#
%
\begin{definition}
%[$(f,\X)$-algorithm]
\label{deffactory}
Let $\X$ be a subset of $\R$ and $f\dvtx \conv(\X) \to\R^+$ a function.
An $f$-factory $\A\equiv(\phi,T)$ is
an $\X$-algorithm such that for any distribution $\pi\in\mathcal
{M}_1(\X)$, an independent sequence $X=(X_n)_{n
\geq1}$ marginally distributed as $\pi$ and an auxiliary random
variable $U \sim\operatorname{Uniform}(0,1)$ independent of $(X_n)_{n \geq1}$,
the random variable $Y=\A(U,X)$ is a nonnegative unbiased estimator of
$f(m_1(\pi))$.
\end{definition}

The condition $\E(\A(U,X)) = f(m_1(\pi))$ implies that the algorithm
terminates with probability one when fed with the independent sequence
$X=(X_n)_{n \geq1}$ and $U \sim
\operatorname{Uniform}(0,1)$. Importantly the definition implies that an
$f$-factory should work for any distribution $\pi\in\mathcal{M}_1(\X)$.

%s2.2 #&#
\subsection{Nonexistence of general $f$-factories}\label{subseccasereal}

We first consider the general case $\X=\R$, where the unbiased
estimators used as input can take any real value.
%th2.1 #&#

\begin{teo}
For any nonconstant function $f\dvtx\R\to\R^+$, no $f$-factory exists.
\label{teocaserealnonexistence}
\end{teo}

\begin{pf}
For the sake of contradiction, suppose that there exists a nonconstant
function $f\dvtx\R\to\R^+$ and an $\R$-algorithm $(\phi,T)$ as in Definition
\ref{deffactory}; because $f$ is not constant, there exist two real numbers
$\lambda_X,\lambda_Y \in\R$ with $f(\lambda_X)>f(\lambda_Y)$.
Choose any distribution $\mu_X \in\M_1(\R)$ with $m_1(\mu_X) =
\lambda_X$, and
consider a sequence $X=(X_n)_{n\geq1}$ marginally distributed
according to
$\mu_X$. For $\varepsilon>0$ and an independent sequence of Bernoulli random
variables $(B_n)_{n \geq1}$ with success probability
$\p(B_n=1)=1-\p(B_n=0)=1-\varepsilon$,
independent from any other source of randomness, the sequence
$Y=(Y_n)_{n\geq
1}$ defined by
%
%e2.1 #&#
%
\begin{equation}
\label{eqcoupling} Y_n = B_n X_n +
\frac{\lambda_Y-\lambda_X(1-\varepsilon
)}{\varepsilon} (1-B_n)
\end{equation}
is such that $\E(Y_n)=\lambda_Y$. For any integer $n$ we have
$Y_n=X_n$ with arbitrarily large probability $1-\varepsilon$, where
$\varepsilon$ can be chosen arbitrarily small, while $\lambda_Y$
and $\lambda_X$ are distinct and fixed; this construction is pivotal
in all the proofs of this article.

%In layman's terms
Let us first give an informal description of the proof.
We will compare the outputs of the algorithm
for the two input sequences $(X_n)_{n\geq1}$ and $(Y_n)_{n\geq1}$ and
a common auxiliary variable $U$.
Suppose first that the algorithm terminates after $n$ steps when fed
with the
sequence $(X_n)_{n\geq1}$.
By tuning the value of $\varepsilon$ we can make the events
$\{(Y_1, \ldots, Y_n) \neq(X_1, \ldots, X_n)\}$ arbitrarily rare.
On the other hand the expected outputs are set to $f(\lambda_X)$ for
$(X_n)_{n\geq1}$ and $f(\lambda_Y)$ for $(Y_n)_{n\geq1}$,
with $f(\lambda_Y) < f(\lambda_X)$.
Hence, when the rare events $\{(Y_1, \ldots, Y_n) \neq(X_1, \ldots,
X_n)\}$ do occur,
the algorithm using $(Y_1, \ldots, Y_n)$ needs to output a value
sufficiently smaller than the value produced by the algorithm using
$(X_1, \ldots, X_n)$, so that
the expected output can shift from $f(\lambda_X)$ to $f(\lambda_Y)$.
However, the algorithm is not allowed to produce negative values so
that the minimum output is zero. This would lead to a contradiction
when the events $\{(Y_1, \ldots, Y_n) \neq(X_1, \ldots, X_n)\}$ are
rare enough.

More formally denote by $\mu_Y$ the marginal law of each $Y_n$, namely
\[
\mu_Y(dy) = (1-\varepsilon)\mu_X(dy) +
\varepsilon1_{\varepsilon
^{-1} (\lambda_Y-\lambda_X(1-\varepsilon))}(dy).
\]
The joint law on $([0,1], \R^\N, \R^\N)$ of the random variables $(U,
(X_n)_{n\geq1}, (Y_n)_{n\geq1})$ is denoted by $\check\mu$; the
marginal of
$\check\mu$ on its first two arguments is $(\operatorname{Uniform}(0,1),\break 
\mu_X^{\otimes\N})$, and the marginal on its first and third
arguments is
$(\operatorname{Uniform}(0,1), \mu_Y^{\otimes\N})$. We denote by $\check
\E$ the
expectation with respect to $\check\mu$ and by $\E_{U,X}$ and $\E
_{U,Y}$ the
expectations with respect to those two marginals, respectively.

Recall that the stopping times
\begin{eqnarray*}
\tau_X &=& \inf\bigl\{n\dvtx T_n(U, X_1,
\ldots, X_n) = 1 \bigr\}, \qquad\tau_Y = \inf\bigl\{n\dvtx
T_n(U, Y_1, \ldots, Y_n) = 1 \bigr\}
\end{eqnarray*}
are by assumption almost surely finite and
\[
\E_{U,X}\bigl(\phi_{\tau_X}(U,X_1, \ldots,
X_{\tau_X})\bigr) = f(\lambda_X),\qquad\E_{U,Y}
\bigl(\phi_{\tau_Y}(U,Y_1, \ldots, Y_{\tau_Y})\bigr) = f(
\lambda_Y).
\]
Notice further that
\[
\tau_X 1_{L_n\cap M_n} = \tau_Y 1_{L_n \cap M_n},
\]
where we have defined the sets
$L_n = \{ \omega\dvtx \tau_X \leq n \}$ and
$M_n = \{ \omega\dvtx B_1=\cdots=B_n=1 \} \subseteq\{\omega\dvtx X_1
= Y_1,
\ldots, X_n=Y_n\}$.
Since $\phi_{\tau_Y}$ is almost surely nonnegative, we have for all
$n\geq1$,
%e2.2 #&#
%
\begin{eqnarray}\label{eqnonneg}
\nonumber
\E_{U,Y} \bigl( \phi_{\tau_Y}(U, Y_1,
\ldots, Y_{\tau_Y}) \bigr) &=& \check\E\bigl( \phi_{\tau_Y}(U,
Y_1, \ldots, Y_{\tau
_Y}) \bigr)
\\
&\geq&\check\E\bigl( \phi_{\tau_Y}(U, Y_1, \ldots,
Y_{\tau_Y}) 1_{L_n\cap M_n} \bigr)
\\
&=& \check\E\bigl( \phi_{\tau_X}(U, X_1, \ldots,
X_{\tau_X}) 1_{L_n\cap M_n} \bigr).\nonumber
\end{eqnarray}
The random variables $(B_n)_{n \geq1}$ are independent of any other
source of randomness so that for all $n\geq1$, we have
%e2.3 #&#
%
\begin{eqnarray}\label{eqbernoulli}
&& \check\E\bigl( \phi_{\tau_X}(U, X_1, \ldots,
X_{\tau
_X}) 1_{L_n\cap M_n} \bigr)\nonumber
\\
&&\qquad = (1-\varepsilon)^n \check
\E\bigl( \phi_{\tau_X}(U, X_1, \ldots, X_{\tau_X})
1_{L_n} \bigr)
\\
&&\qquad =  (1-\varepsilon)^n \E_{U,X} \bigl( \phi_{\tau_X}(U,
X_1, \ldots, X_{\tau_X}) 1_{L_n} \bigr).\nonumber
\end{eqnarray}
The dominated convergence theorem yields
\begin{eqnarray*}
\lim_{n \to\infty} \E_{U,X} \bigl( \phi_{\tau_X}(U,
X_1, \ldots, X_{\tau_X}) 1_{L_n} \bigr) &=&
\E_{U,X} \bigl( \phi_{\tau_X}(U, X_1, \ldots,
X_{\tau_X}) \bigr)
\\
&=& f(\lambda_X)
\end{eqnarray*}
so that for any $\delta>0$, there exists $n_0 = n_0(\delta) \in\N$
such that for all $n\geq n_0$,
%
%e2.4 #&#
%
\begin{equation}
f(\lambda_X)-\delta\leq\E_{U,X} \bigl(
\phi_{\tau_X}(U, X_1, \ldots, X_{\tau_X})
1_{L_n} \bigr) \leq f(\lambda_X).\label{eqlimit}
\end{equation}
One can choose $\delta> 0$ and $\eta> 0$ such that $f(\lambda_Y) +
\eta< f(\lambda_X) - \delta$.
Equations (\ref{eqnonneg}), (\ref{eqbernoulli}) and (\ref
{eqlimit}) yield that
for some integer $n_0 = n_0(\delta)$ and any $\varepsilon>0$, we have
\begin{eqnarray*}
f(\lambda_Y) &=& \E_{U,Y} \bigl( \phi_{\tau_Y}(U,
Y_1, \ldots, Y_{\tau_Y}) \bigr) \geq\check\E\bigl(
\phi_{\tau_X}(U, X_1, \ldots, X_{\tau_X})
1_{L_{n_0}\cap M_{n_0}} \bigr)
\\
&=& (1-\varepsilon)^{n_0} \E_{U,X} \bigl( \phi_{\tau_X}(U,
X_1, \ldots, X_{\tau_X}) 1_{L_{n_0}} \bigr)
\\
&\geq& (1 - \varepsilon)^{n_0} \bigl(f(\lambda_X)-\delta
\bigr) > (1-\varepsilon)^{n_0}\bigl(f(\lambda_Y) + \eta
\bigr).
\end{eqnarray*}
We obtain a contradiction for $\varepsilon> 0$ small enough.
\end{pf}

Theorem \ref{teocaserealnonexistence} indicates in particular that
given $U$-estimators $(X_n)_{n \geq1}$ of a quantity
$\lambda$ and without additional knowledge on these estimators, we
cannot obtain $U^+$-estimators of
neither $\exp(\lambda)$ nor $1/\lambda$.

Another question of interest arises in the case where $\mathcal{X} =
\R$, we
are given \mbox{$U$-}estimators of a quantity $\lambda>0$ and we want to
construct a
$U^+$-estimator $Y$ of the same quantity $\lambda$. This is not exactly
equivalent to asking whether there exists an $f$-factory for
$f\dvtx x\mapsto x$,
first because we have only defined $f$-factories for $f$ taking values in
$\R^+$, and second because in Definition \ref{deffactory} the algorithm
should work for any variable distributed as $\pi\in\mathcal{M}_1(\R)$,
whereas here we only consider distributions with expectation in $\R^+$.

%le2.1 #&#
%
\begin{lemma}\label{lemmaRisnotpositivable}
Let $\eta\geq0$ be a known constant.
There does not exist an \mbox{$\R$-}algorithm $\A\equiv(\phi,T)$ such that
for any
independent sequence $X=(X_n)_{n \geq1}$ marginally distributed as
$\pi\in
\mathcal{M}_1(\R)$ with $m_1(\pi)> \eta$ and an auxiliary random
variable $U \sim
\operatorname{Uniform}(0,1)$ independent from $(X_n)_{n \geq1}$, the random
variable $Y=\A(U,X)$ is a nonnegative unbiased estimator of $m_1(\pi)$.
\end{lemma}

\begin{pf}
We follow the same arguments as in the proof of Theorem \ref{teocaserealnonexistence}.
Consider $\lambda_X,\lambda_Y \in\R^+$ with $\lambda_X > \lambda
_Y > \eta$,
and an algorithm $\A\equiv(\phi,T)$ as in the statement of Lemma~\ref{lemmaRisnotpositivable}.
Let $\mu_X \in\M_1(\R)$ with $m_1(\mu_X) = \lambda_X$, and
consider an
sequence $X=(X_n)_{n\geq1}$ marginally distributed according to $\mu_X$.
One can define $Y$ as in equation (\ref{eqcoupling}). Since $\E(Y) =
\lambda_Y \geq0$, one can construct the same contradiction as in the proof
of Theorem \ref{teocaserealnonexistence}.
\end{pf}

The presence of $\eta\geq0$ in the statement might seem cumbersome but
emphasizes that the contradiction does not stem from distributions with
expectation arbitrarily close to zero. According to the Lemma~\ref{lemmaRisnotpositivable}, even if one knows that a sequence of estimators
has expectation larger than one, say, it is still impossible to design
an algorithm
transforming that sequence into a nonnegative random variable with the same
expectation.

In the light of the nonexistence of $f$-factories when $\X= \R$, as
stated in Theorem \ref{teocaserealnonexistence},
we propose to study their existence when $\X$ is a subset of $\R$ in
the next section.

%
%s3 #&#
\section{Existence under stronger assumptions}\label{secexistencecases}

%
%s3.1 #&#
\subsection{Case where \texorpdfstring{$\mathcal{X}=[a,+\infty)$}{X=[a,+infty)} or 
\texorpdfstring{$\mathcal{X}=(-\infty,b]$}{X=(-infty,b]}}\label{subseccaserealplus}

%le3.1 #&#
%
\begin{lemma} \label{lemmacaserealplus} Let $a,b \in\R$ be two real numbers:
\begin{itemize}
\item For an $f$-factory to exist with $\X=[a,\infty)$ and $f\dvtx\X\to
\R^+$, $f$ must be increasing.
\item For a $g$-factory to exist with $\X=(-\infty,b]$ and $g\dvtx\X\to
\R^+$, $g$ must be decreasing.
\end{itemize}
\end{lemma}

\begin{pf}
By symmetry we prove only the first assertion.
For the sake of contradiction assume that there exist $a \leq\lambda
_X<\lambda_Y$ with
$f(\lambda_X)>f(\lambda_Y)$ and an algorithm $\A\equiv(\phi,T)$ as in
Definition \ref{deffactory}. Choose any distribution $\mu_X \in
\M_1 ([a,\infty) )$ with $m_1(\mu_X) = \lambda_X$ and an
independent sequence
$X=(X_n)_{n\geq1}$ marginally distributed according to $\mu_X$. For
$\varepsilon\in(0,1)$, consider the sequence $Y=(Y_n)_{n\geq1}$ as defined
in equation (\ref{eqcoupling}). For $\varepsilon>0$ small enough we
have $\D(Y) \in\M_1( [a,\infty) )$ since $\lambda_Y>\lambda_X$.
One can then construct exactly the same contradiction as in the proof
of Theorem \ref{teocaserealnonexistence}.
\end{pf}

Lemma \ref{lemmacaserealplus} indicates in particular that it is
impossible to
obtain $U^+$-estimators of $1/\lambda$ given
$U^+$-estimators of a quantity $\lambda>0$ without exploiting any other
additional information on the distribution of these $U^+$-estimators.
For $\X=[a,\infty)$ and some increasing functions $f$, there can be
explicit constructions of $f$-factories.
For example, there exists an $f$-factory for any function $f\dvtx
[a,\infty
) \to
\R^+$ that can be expressed as a power series of the type
%
%e3.1 #&#
%
\begin{equation}
\label{eqfanalytic} f(x) = \sum_{n=0}^{\infty}
c_n (x-a)^n\qquad\mbox{with } c_n\geq0
\mbox{ for all } n\geq0.
\end{equation}
Indeed, introduce an independent sequence of random variables
$(X_n)_{n\geq1}$ margi\-nally distributed as
$\mu_X \in\M_1 ([a,\infty) )$ and an integer-valued random
variable $N$; setting the weights $w_n = 1/\p(N\geq n)$ as in
Section~\ref{subsecdebiasing}, Tonelli's theorem yields that the estimator
\[
Y = \sum_{n=0}^N w_n
c_n \prod_{k=1}^n
(X_{k}-a),
\]
where the product is equal to $1$ when $n=0$,
is well defined, is almost surely nonnegative and has expectation
$f ( m_1(\mu_X) )$.

%
% Example: Poisson estimator
%
The above discussion gives a construction of a \textit{Poisson estimator},
that is, a \mbox{$U^+$-}estimator of $\lambda= \exp(\E[X])$ given a stream
$(X_n)_{n
\geq1}$ of i.i.d. $[a,+\infty)$-valued random variables distributed as $X$.
Indeed the exponential function can be expressed as in equation
(\ref{eqfanalytic}) with $c_n = \exp(a)/n!$.
One can readily check that if $X$ has a finite variance and if the
random variable $N$ does not decay too rapidly to zero, for instance,
$\p(N
\geq n) \geq C / (1+\varepsilon)^n$ for some constants $C,\varepsilon
>0$ as is the case for a geometric random variable, then equation (\ref
{eqvariance}) holds with
\[
S_n = \exp(a)+\sum_{k=1}^n
\frac{\exp(a)}{k!} \prod_{j=1}^k
(X_{j} - a)
\]
and $S=S_{\infty}$.
The resulting Poisson estimator is unbiased and has a finite variance.

%
% Conjecture
%
For increasing functions in general, the existence of $f$-factories
remains an open question.
Denoting by $\mathcal{F}$ the class of functions of the form described
by equation~(\ref{eqfanalytic}), and by $\mathcal{C}$ the class of functions
$f\dvtx[a,+\infty) \to
\mathbb{R}^+$ for which an \mbox{$f$-}factory exists, the previous
discussion shows that $\mathcal{F}\subset\mathcal{C}$, and
we conjecture $\mathcal{F} = \mathcal{C}$.
For $f$ and $g$ in $\mathcal{C}$, then $f+g$ and $f\times g$ are in
$\mathcal{C}$.
In the special case $a = 0$, then $f\circ g$ is also in $\mathcal{C}$.
A random truncation argument also shows that if $h\dvtx[a,+\infty) \to
\mathbb{R}^+$ can be expressed as the infinite sum $h = \sum_{k \geq0}
f_k$ for functions $f_k \in\mathcal{C}$, then $h \in\mathcal{C}$.
The set of functions $\mathcal{F}$ is the smallest class of functions
that contains positive constants and the function $x \mapsto(x-a)$ and
that is
stable by the above-described operations.
Those operations leave $\mathcal{C}$ stable
because of simple properties of the expectation, such as linearity and
the identity $\mathbb{E}[X \times Y] = \mathbb{E}[X] \times\mathbb
{E}[Y]$ for $X$ independent from $Y$.
Our conjecture is based on our inability to exploit other properties of
the expectation
to find functions that would be in $\mathcal{C}$ but not in $\mathcal{F}$.

%s3.2 #&#
\subsection{Case where \texorpdfstring{$\mathcal{X}=[a,b]$}{X=[a,b]}}\label{subseccaseinterval}
The case of a bounded interval $\X=[a,b]$ is the most related to the Bernoulli
factory described in Section~\ref{subsecdebiasing}. We highlight in
this section the similarities and differences
between the construction of nonnegative estimators and Bernoulli factories.
We then give a complete characterization of functions
$f\dvtx\X=[a,b] \to\R^+$ for which $f$-factories exist.

Arguments similar to the proof of Theorem~\ref{teocaserealnonexistence} show that for an $f$-factory to exist,
the function
$f\dvtx\X\to\R^+$ has to be continuous. Such a function $f\dvtx\X\to\R^+$
is thus
necessarily bounded, and we consider a nontrivial interval $[0,\gamma]$
containing its range. If a Bernoulli factory exists for the function
$g\dvtx[0,1] \to[0,1]$ with $g(x) = f(a (1-x) + bx) / \gamma$,
then there exists an $f$-factory. Indeed, consider an i.i.d. sequence
$X=(X_n)_{n\geq1}$
marginally distributed according to $\mu_X \in\M_1(\X)$. Introduce
random variables
$(B_n)_{n\geq1}$, with $B_n:= 1_{U_n \leq(X_n - a)/(b-a)}$ where
$(U_n)_{n\geq1}$ is an i.i.d. sequence of random variables uniformly
distributed on $(0,1)$. Then $(B_n)_{n\geq1}$ forms an i.i.d. sequence
of Bernoulli random variables
with mean $(m_1(\mu_X) - a)/(b-a)$.
Therefore the Bernoulli factory for $g$ takes
the sequence $(B_n)_{n \geq1}$ as input and produces a Bernoulli random
variable $\widetilde{B}$ with mean $g((m_1(\mu_X) -
a)/(b-a))=f(m_1(\mu_X)) / \gamma$. The random
variable $\gamma \widetilde{B}$ is thus a nonnegative unbiased
estimator of
$m_1(\mu_X)$. As proved in \citet{Keane1994}, a necessary and sufficient
condition on $g\dvtx[0,1] \to[0,1]$ for the existence of a Bernoulli factory
is
\[
\exists\varepsilon> 0, \exists n \in\mathbb{N}, \forall x\in
[0,1]\qquad
\min\bigl(g(x), 1-g(x) \bigr) \geq\varepsilon\min\bigl( x^n,
(1-x)^n \bigr).
\]
It
follows that an $f$-factory exists as soon as the condition $\min(f(x),
\gamma-f(x)) \geq\varepsilon \min( (x-a)^n, (b-x)^n
)$
is satisfied for some $\varepsilon> 0$, $n \in\mathbb{N}$ and all
$x\in[a,b]$.
Theorem \ref{teocaseinterval} shows in fact that
%
%e3.2 #&#
%
\begin{equation}\label{eqconditioncnstructibility}
\exists\varepsilon> 0, \exists n \in\mathbb{N}, \forall
x\in[a,b]\qquad f(x) \geq\varepsilon\min\bigl( (x-a)^n,
(b-x)^n \bigr)
\end{equation}
is a necessary and sufficient condition for an $f$-factory to exist. The
necessary condition $1-g(x) \geq\varepsilon \min( x^n,
(1-x)^n
)$ for the Bernoulli factory problem to have a solution comes from
the fact
that the Bernoulli factory has to produce a $\{0,1\}$-valued estimator;
we only
need to construct a $[0,\infty)$-valued estimator and can thus get
away with
the weaker condition (\ref{eqconditioncnstructibility}).

%
% THM: interval case
%
%th3.1 #&#
%
\begin{teo} \label{teocaseinterval}
Let $\X=[a,b]$ be a real interval and $f\dvtx\X\to\R^+$ a continuous
function that is not identically zero.
There exists an $f$-factory if and only if
condition (\ref{eqconditioncnstructibility}) holds.
\end{teo}

\begin{pf} The sufficiency is proved as a consequence of the results
proved in \citet{Keane1994}. The proof of the necessity requires
different arguments.

\begin{longlist}
\item[\textit{Sufficiency}.] Let $f\dvtx\X\to\R^+$ be a continuous function that satisfies condition~(\ref{eqconditioncnstructibility}). Since $f$ is bounded on $\X$,
one can
find $\gamma\geq\max_{x \in\X} f(x)$ large enough such that
$\gamma-f(x) > \varepsilon \min( (x-a)^n, (b-x)^n
)$ for all $x \in\X$. The
discussion before the statement of Theorem \ref{teocaseinterval} thus
shows that an $f$-factory can be constructed.
\end{longlist}

\begin{longlist}
\item[\textit{Necessity}.]
For notational convenience, we present the proof in the case $\X
=[0,1]$. The
general case $\X=[a,b]$ is identical. Let $\A\equiv(T,\phi)$ be an
$f$-factory for some function $f\dvtx[0,1] \to\R^+$. For $x_{1\dvtx n}
= (x_1, \ldots,
x_n) \in
\{0,1\}^n$
and a random variable $U$ uniformly distributed on $(0,1)$,
we denote by $F_n(x_{1\dvtx n})$ the set of events such that the algorithm
terminates after having processed $x_{1\dvtx n}$, that is,
\begin{eqnarray*}
F_n(x_{1\dvtx n}) &=& \bigl\{\omega\dvtx \inf\bigl\{ 1 \leq k \leq
n\dvtx
T_k(U, x_1, \ldots, x_k) = 1\bigr\} = n
\bigr\}
\end{eqnarray*}
with the convention $\inf\{ \varnothing\} = \infty$.
We define the expected output given $x_{1\dvtx n}$ by
\begin{eqnarray*}
\Psi_n(x_{1\dvtx n}) &=& \E\bigl( 1_{F_n(x_{1\dvtx n})}
\phi_n(U, x_1, \ldots, x_n) \bigr).
\end{eqnarray*}
For any index $n \geq1$ and $x_{1\dvtx n} \in\{0,1\}^n$,
$\Psi_n(x_{1\dvtx n})$ is a nonnegative real number. By Definition
\ref{deffactory} for any $z \in[0,1]$
and an i.i.d. sequence $(X_n)_{n \geq1}$ of Bernoulli random variables
with mean $z \in[0,1]$,
we have
\begin{eqnarray*}
f(z) &=& \E\Biggl( \sum_{n=1}^\infty
\Psi_n(X_{1\dvtx n}) \Biggr) = \sum_{n=1}^\infty
\sum_{x_{1\dvtx n} \in\{0,1\}^n } \p(X_{1\dvtx n} = x_{1\dvtx n})
\Psi_n(x_{1\dvtx n}).
\end{eqnarray*}
For any index $n \geq1$ and $x_{1\dvtx n} \in\{0,1\}^n$,
defining $r=r(x_{1\dvtx n}) = \operatorname{Card}\{1\leq i \leq n\dvtx
x_i=1\}$, we
have $\p(X_{1\dvtx n} = x_{1\dvtx n}) = z^r (1-z)^{n-r}$,
and the above double sum can be written as
\begin{eqnarray*}
f(z) &=& \sum_{n=1}^\infty\sum
_{x_{1\dvtx n} \in\{0,1\}^n } z^r (1-z)^{n-r}
\Psi_n(x_{1\dvtx n}) = \sum_{p,q \in\mathbb{N}^2}
c_{p,q} z^p (1-z)^q
\end{eqnarray*}
for some nonnegative coefficient $c_{p,q} \geq0$. Condition (\ref
{eqconditioncnstructibility}) follows.\quad\qed
\end{longlist}\noqed
\end{pf}

By Theorem \ref{teocaseinterval} it is possible to obtain
$U^+$-estimators of
$\exp(\lambda)$ or $1/\lambda$ given $U$-estimators of $\lambda$
with support in
some known interval $[a,b]$. Indeed, for the exponential case, one can use
either a Bernoulli factory or the Poisson estimator described at the
end of
Section~\ref{subseccaserealplus}. For the inverse case on a segment $[a,b]
\subset(0, \infty)$, one can use either a Bernoulli factory or a random
truncation argument to the series expansion
\[
\frac{1}{x} = \frac{1}{b} \sum_{k=0}^{\infty}
\biggl( \frac
{b-x}{b} \biggr)^k
\]
to construct an unbiased estimate of $\lambda= 1 / \E[X]$ given a
stream $(X_n)_{n \geq1}$ of i.i.d. $[a,b]$-valued random variables
distributed as $X$.

%
%s4 #&#
\section{Discussion}\label{secdiscussion}

%s4.1 #&#
\subsection{Summary of the analysis}\label{subseclimits}

The results of Section~\ref{subseccasereal} show that, for a nonconstant
function $f\dvtx\R\to\R^+$, the ability to sample an unbiased estimator\vadjust{\goodbreak}
$X$ of a
quantity $\lambda$ is not enough to obtain a nonnegative unbiased
estimator of
$f(\lambda)$. However, as described in Section~\ref
{secexistencecases}, when
additional information such as almost sure lower or upper bounds on $X$ is
available, an $f$-factory might exist.
The case where $f$ is
increasing and the support of $X$ is $[a,\infty)$ remains partly unsettled.

We have prescribed as input of $f$-factories unbiased estimators of arbitrary
quantities $\lambda\in\R$; other types of input could be envisioned,
such as
estimators consistent in $L_2$. However, in this
case we could first apply a debiasing technique recalled in
Section~\ref{subsecdebiasing} and then feed the output to an
$f$-factory, and hence the conclusion would be similar.
Finally we have not considered the multi-dimensional case $f\dvtx\R^{d}
\to\R^{+}$ for $d
> 1$ since, in the context of exact inference, quantities of interest are
posterior density evaluations.

%s4.2 #&#
\subsection{Exact or inexact inference}

An advantage of exact methods, where no systematic bias remains, is
that the
error is entirely due to the variation in the Monte Carlo algorithm and
thus is
straightforward to quantify and to interpret [\citet{wagner1987unbiased}].
The trade-off between computational feasibility and exactness is
ubiquitous in
statistics, for instance, between Ensemble Kalman filters and particle filters
[\citet{frei2013bridging}] or between approximate Bayesian
computation and Markov
chain Monte Carlo [\citet{marin2012approximate}].
In some contexts such as state space models, a
nonnegative unbiased estimator of the likelihood can be directly obtained,
and the pseudo-marginal approach is proven efficient
[\citet{andrieudoucetholenstein2010}].
Our study indicates that in some contexts nonnegative unbiased estimators
cannot be obtained, and thus the pseudo-marginal approach cannot be applied.
Exact inference could still be performed using signed unbiased estimators,
as in the computational physics literature
[\citet{lin2000noisy,troyer2005computational,Girolami2013}].

In Section~\ref{secexistencecases}
the existence of $f$-factories has been studied under additional
assumptions on the support
of the input sequence. These assumptions are consistent with recent
Monte Carlo methods for large datasets
that take advantage of almost sure bounds to bypass the evaluation of
the full likelihood
[\citet{bardenet2014,maclaurin2014firefly}], leading to exact
methods or inexact methods with a
controlled error.
There exist inexact methods with no control of the bias, which do not
require almost sure bounds,
such as some approximations
of Metropolis--Hastings algorithms [\citet
{ceperley1999penalty,nicholls2012coupled}]
or of Langevin diffusions [\citet
{welling2011bayesian,ahn2012bayesian,chen2014stochastic}].

When $f$-factories exist as in Section~\ref{secexistencecases}, we have
discussed implementable schemes based on the Bernoulli factory or on
random truncations
of infinite series. The algorithms considered in Definition \ref
{deffactory} terminate with
probability one, but the expected computational time is not necessarily finite.
Hence even if the method could be applied in principle, its computational
cost might prevent any practical implementation.
The recent literature on Bernoulli factories has
focused on characterizing algorithms that generate the desired output
using as few
input variables as possible
[\citet
{Nacu2005,latuszynski2011simulating,thomas2011practical,flegal2012exact}],
whereas \citet{RheeGlynn2012,RheeGlynn2013} carefully study
the expected computational
cost of debiasing techniques.
The minimum computational cost of $f$-factories could be studied as well.

\section*{Acknowledgments}
The first author gratefully acknowledges EPSRC for funding this research
through grant EP/K009362/1. We are grateful to Mourad Sabilellah for stimulating
discussions. We
thank the Associate Editor and two anonymous referees for their
comments, which
helped improving both the presentation and the content of this article.


%suskaldyti doi

% imsref loaded by linak, 2015-02-19 15:52:52
%
% imsref loaded by linak, 2015-02-20 15:47:58

\printaddresses

\begin{thebibliography}{46}

%b1 ###
\bibitem[\protect\citeauthoryear{Ahn, Korattikara and Welling}{2012}]{ahn2012bayesian}
%
\begin{binproceedings}[author]
\bauthor{\bsnm{Ahn},~\bfnm{Sungjin}\binits{S.}},
\bauthor{\bsnm{Korattikara},~\bfnm{Anoop}\binits{A.}} \AND
\bauthor{\bsnm{Welling},~\bfnm{Max}\binits{M.}}
(\byear{2012}).
\btitle{Bayesian posterior sampling via stochastic gradient Fisher scoring}.
In \bbooktitle{Proceedings of the 29th International Conference on
Machine Learning (ICML-12)}
\bpages{1591--1598}.
\end{binproceedings}
%
%\OrigBibText
%%
%\begin{binproceedings}[author]
%\bauthor{\bsnm{Ahn},~\bfnm{Sungjin}\binits{S.}},
%\bauthor{\bsnm{Korattikara},~\bfnm{Anoop}\binits{A.}} \AND
%\bauthor{\bsnm{Welling},~\bfnm{Max}\binits{M.}}
%(\byear{2012}).
%\btitle{Bayesian posterior sampling via stochastic gradient Fisher scoring}.
%In \bbooktitle{Proceedings of the 29th International Conference on Machine
%Learning (ICML-12)}.
%\end{binproceedings}
%%
%\endOrigBibText
\bptok{imsref}%
\endbibitem

%b2 ###
\bibitem[\protect\citeauthoryear{Andrieu, Doucet and
Holenstein}{2010}]{andrieudoucetholenstein2010}
%
\begin{barticle}[author]
\bauthor{\bsnm{Andrieu},~\bfnm{C.}\binits{C.}},
\bauthor{\bsnm{Doucet},~\bfnm{A.}\binits{A.}} \AND
\bauthor{\bsnm{Holenstein},~\bfnm{R.}\binits{R.}}
(\byear{2010}).
\btitle{Particle {M}arkov chain {M}onte {C}arlo (with discussion)}.
\bjournal{J. Roy. Statist. Soc. Ser. B}
\bvolume{72}
\bpages{357--385}.
\end{barticle}
%
%\OrigBibText
%%
%\begin{barticle}[author]
%\bauthor{\bsnm{Andrieu},~\bfnm{C.}\binits{C.}},
%\bauthor{\bsnm{Doucet},~\bfnm{A.}\binits{A.}} \AND
%\bauthor{\bsnm{Holenstein},~\bfnm{R.}\binits{R.}}
%(\byear{2010}).
%\btitle{Particle {M}arkov chain {M}onte {C}arlo (with discussion)}.
%\bjournal{J. Royal Statist. Society Series B}
%\bvolume{72}
%\bpages{357--385}.
%\end{barticle}
%%
%\endOrigBibText
\bptok{imsref}%
\endbibitem

%b3 ###
\bibitem[\protect\citeauthoryear{Andrieu and
Roberts}{2009}]{andrieuroberts2009}
%
\begin{barticle}[mr]
\bauthor{\bsnm{Andrieu},~\bfnm{Christophe}\binits{C.}} \AND
\bauthor{\bsnm{Roberts},~\bfnm{Gareth~O.}\binits{G.~O.}}
(\byear{2009}).
\btitle{The pseudo-marginal approach for efficient {M}onte {C}arlo
computations}.
\bjournal{Ann. Statist.}
\bvolume{37}
\bpages{697--725}.
\bid{doi={10.1214/07-AOS574}, issn={0090-5364}, mr={2502648}}
\end{barticle}
%
%\OrigBibText
%%
%\begin{barticle}[author]
%\bauthor{\bsnm{Andrieu},~\bfnm{C.}\binits{C.}} \AND
%\bauthor{\bsnm{Roberts},~\bfnm{G.~O.}\binits{G.~O.}}
%(\byear{2009}).
%\btitle{{The pseudo-marginal approach for efficient {M}onte {C}arlo
%computations.}}
%\bjournal{Ann. Statist.}
%\bvolume{37}
%\bpages{697-725}.
%\bdoi{10.1214/07-AOS574}
%\end{barticle}
%%
%\endOrigBibText
\bptok{imsref}%
% NOT OUTPUTTED:
% number = 2
% doi = http://dx.doi.org/10.1214/07-AOS574
% coden = ASTSC7
% fjournal = The Annals of Statistics
\endbibitem

%b4 ###
\bibitem[\protect\citeauthoryear{Bardenet, Doucet and Holmes}{2014}]{bardenet2014}
%
\begin{binproceedings}[author]
\bauthor{\bsnm{Bardenet},~\bfnm{R{\'e}mi}\binits{R.}},
\bauthor{\bsnm{Doucet},~\bfnm{Arnaud}\binits{A.}} \AND
\bauthor{\bsnm{Holmes},~\bfnm{Chris}\binits{C.}}
(\byear{2014}).
\btitle{Towards scaling up Markov chain {M}onte {C}arlo: An adaptive
subsampling approach}.
In \bbooktitle{Proceedings of the 31st International Conference on
Machine Learning (ICML-14)}
\bpages{405--413}.
\end{binproceedings}
%
%\OrigBibText
%%
%\begin{binproceedings}[author]
%\bauthor{\bsnm{Bardenet},~\bfnm{R{\'e}mi}\binits{R.}},
%\bauthor{\bsnm{Doucet},~\bfnm{Arnaud}\binits{A.}} \AND
%\bauthor{\bsnm{Holmes},~\bfnm{Chris}\binits{C.}}
%(\byear{2014}).
%\btitle{Towards scaling up Markov chain {M}onte {C}arlo: an adaptive
%subsampling approach}.
%In \bbooktitle{Proceedings of The 31st International Conference on Machine
%Learning}
%\bpages{405--413}.
%\end{binproceedings}
%%
%\endOrigBibText
\bptok{imsref}%
\endbibitem

%b5 ###
\bibitem[\protect\citeauthoryear{Beaumont}{2003}]{beaumont2003}
%
\begin{barticle}[author]
\bauthor{\bsnm{Beaumont},~\bfnm{M.~A.}\binits{M.~A.}}
(\byear{2003}).
\btitle{Estimation of population growth or decline in genetically
monitored populations}.
\bjournal{Genetics}
\bvolume{164}
\bpages{1139--1160}.
\end{barticle}
%
%\OrigBibText
%%
%\begin{barticle}[author]
%\bauthor{\bsnm{Beaumont},~\bfnm{M.~A.}\binits{M.~A.}}
%(\byear{2003}).
%\btitle{Estimation of population growth or decline in genetically monitored
%populations}.
%\bjournal{Genetics}
%\bvolume{164}
%\bpages{1139-1160}.
%\end{barticle}
%%
%\endOrigBibText
\bptok{imsref}%
\endbibitem

%b6 ###
\bibitem[\protect\citeauthoryear{Berger, Bernardo and
Sun}{2009}]{berger2009formal}
%
\begin{barticle}[mr]
\bauthor{\bsnm{Berger},~\bfnm{James~O.}\binits{J.~O.}},
\bauthor{\bsnm{Bernardo},~\bfnm{Jos{\'e}~M.}\binits{J.~M.}} \AND
\bauthor{\bsnm{Sun},~\bfnm{Dongchu}\binits{D.}}
(\byear{2009}).
\btitle{The formal definition of reference priors}.
\bjournal{Ann. Statist.}
\bvolume{37}
\bpages{905--938}.
\bid{doi={10.1214/07-AOS587}, issn={0090-5364}, mr={2502655}}
\end{barticle}
%
%\OrigBibText
%%
%\begin{barticle}[author]
%\bauthor{\bsnm{Berger},~\bfnm{James~O}\binits{J.~O.}},
%\bauthor{\bsnm{Bernardo},~\bfnm{Jos{\'e}~M}\binits{J.~M.}} \AND
%\bauthor{\bsnm{Sun},~\bfnm{Dongchu}\binits{D.}}
%(\byear{2009}).
%\btitle{The formal definition of reference priors}.
%\bjournal{Ann. Statist.}
%\bpages{905--938}.
%\end{barticle}
%%
%\endOrigBibText
\bptok{imsref}%
% NOT OUTPUTTED:
% number = 2
% doi = http://dx.doi.org/10.1214/07-AOS587
% coden = ASTSC7
% fjournal = The Annals of Statistics
\endbibitem

%b7 ###
\bibitem[\protect\citeauthoryear{Beskos, Papaspiliopoulos and
Roberts}{2006}]{beskos2006retrospective}
%
\begin{barticle}[mr]
\bauthor{\bsnm{Beskos},~\bfnm{Alexandros}\binits{A.}},
\bauthor{\bsnm{Papaspiliopoulos},~\bfnm{Omiros}\binits{O.}} \AND
\bauthor{\bsnm{Roberts},~\bfnm{Gareth~O.}\binits{G.~O.}}
(\byear{2006}).
\btitle{Retrospective exact simulation of diffusion sample paths with
applications}.
\bjournal{Bernoulli}
\bvolume{12}
\bpages{1077--1098}.
\bid{doi={10.3150/bj/1165269151}, issn={1350-7265}, mr={2274855}}
\end{barticle}
%
%\OrigBibText
%%
%\begin{barticle}[author]
%\bauthor{\bsnm{Beskos},~\bfnm{Alexandros}\binits{A.}},
%\bauthor{\bsnm{Papaspiliopoulos},~\bfnm{Omiros}\binits{O.}} \AND
%\bauthor{\bsnm{Roberts},~\bfnm{Gareth~O}\binits{G.~O.}}
%(\byear{2006}).
%\btitle{Retrospective exact simulation of diffusion sample paths with
%applications}.
%\bjournal{Bernoulli}
%\bvolume{12}
%\bpages{1077--1098}.
%\end{barticle}
%%
%\endOrigBibText
\bptok{imsref}%
% NOT OUTPUTTED:
% number = 6
% doi = http://dx.doi.org/10.3150/bj/1165269151
% fjournal = Bernoulli. Official Journal of the Bernoulli Society for
%Mathematical Statistics and Probability
\endbibitem

%b8 ###
\bibitem[\protect\citeauthoryear{Beskos and Roberts}{2005}]{beskos2005exact}
%
\begin{barticle}[mr]
\bauthor{\bsnm{Beskos},~\bfnm{Alexandros}\binits{A.}} \AND
\bauthor{\bsnm{Roberts},~\bfnm{Gareth~O.}\binits{G.~O.}}
(\byear{2005}).
\btitle{Exact simulation of diffusions}.
\bjournal{Ann. Appl. Probab.}
\bvolume{15}
\bpages{2422--2444}.
\bid{doi={10.1214/105051605000000485}, issn={1050-5164}, mr={2187299}}
\end{barticle}
%
%\OrigBibText
%%
%\begin{barticle}[author]
%\bauthor{\bsnm{Beskos},~\bfnm{Alexandros}\binits{A.}} \AND
%\bauthor{\bsnm{Roberts},~\bfnm{Gareth~O}\binits{G.~O.}}
%(\byear{2005}).
%\btitle{Exact simulation of diffusions}.
%\bjournal{Ann. Applied Probability}
%\bvolume{15}
%\bpages{2422--2444}.
%\end{barticle}
%%
%\endOrigBibText
\bptok{imsref}%
% NOT OUTPUTTED:
% number = 4
% doi = http://dx.doi.org/10.1214/105051605000000485
% fjournal = The Annals of Applied Probability
\endbibitem

%b9 ###
\bibitem[\protect\citeauthoryear{Beskos et~al.}{2006}]{beskos2006exact}
%
\begin{barticle}[mr]
\bauthor{\bsnm{Beskos},~\bfnm{Alexandros}\binits{A.}},
\bauthor{\bsnm{Papaspiliopoulos},~\bfnm{Omiros}\binits{O.}},
\bauthor{\bsnm{Roberts},~\bfnm{Gareth~O.}\binits{G.~O.}} \AND
\bauthor{\bsnm{Fearnhead},~\bfnm{Paul}\binits{P.}}
(\byear{2006}).
\btitle{Exact and computationally efficient likelihood-based
estimation for discretely observed diffusion processes (with discussion)}.
\bjournal{J. R. Stat. Soc. Ser. B. Stat. Methodol.}
\bvolume{68}
\bpages{333--382}.
\bid{doi={10.1111/j.1467-9868.2006.00552.x}, issn={1369-7412}, mr={2278331}}
\bptnote{check related}%
\end{barticle}
%%
%\OrigBibText
%%
%\begin{barticle}[author]
%\bauthor{\bsnm{Beskos},~\bfnm{Alexandros}\binits{A.}},
%\bauthor{\bsnm{Papaspiliopoulos},~\bfnm{Omiros}\binits{O.}},
%\bauthor{\bsnm{Roberts},~\bfnm{Gareth~O}\binits{G.~O.}} \AND
%\bauthor{\bsnm{Fearnhead},~\bfnm{Paul}\binits{P.}}
%(\byear{2006}).
%\btitle{Exact and computationally efficient likelihood-based
%estimation for
%discretely observed diffusion processes (with discussion)}.
%\bjournal{J. Royal Statist. Society Series B}
%\bvolume{68}
%\bpages{333--382}.
%\end{barticle}
%%
%\endOrigBibText
\bptok{imsref}%
% NOT OUTPUTTED:
% number = 3
% doi = http://dx.doi.org/10.1111/j.1467-9868.2006.00552.x
% fjournal = Journal of the Royal Statistical Society. Series B.
%Statistical Methodology
\endbibitem

%b10 ###
\bibitem[\protect\citeauthoryear{Bhanot and
Kennedy}{1985}]{bhanot1985bosonic}
%
\begin{barticle}[author]
\bauthor{\bsnm{Bhanot},~\bfnm{Gyan}\binits{G.}} \AND
\bauthor{\bsnm{Kennedy},~\bfnm{A.~D.}\binits{A.~D.}}
(\byear{1985}).
\btitle{Bosonic lattice gauge theory with noise}.
\bjournal{Phys. Lett. B}
\bvolume{157}
\bpages{70--76}.
\end{barticle}
%
%\OrigBibText
%%
%\begin{barticle}[author]
%\bauthor{\bsnm{Bhanot},~\bfnm{Gyan}\binits{G.}} \AND
%\bauthor{\bsnm{Kennedy},~\bfnm{AD}\binits{A.}}
%(\byear{1985}).
%\btitle{Bosonic lattice gauge theory with noise}.
%\bjournal{Physics Letters B}
%\bvolume{157}
%\bpages{70--76}.
%\end{barticle}
%%
%\endOrigBibText
\bptok{imsref}%
\endbibitem

%b11 ###
\bibitem[\protect\citeauthoryear{Ceperley and
Dewing}{1999}]{ceperley1999penalty}
%
\begin{barticle}[author]
\bauthor{\bsnm{Ceperley},~\bfnm{D.~M.}\binits{D.~M.}} \AND
\bauthor{\bsnm{Dewing},~\bfnm{M.}\binits{M.}}
(\byear{1999}).
\btitle{The penalty method for random walks with uncertain energies}.
\bjournal{J. Chem. Phys.}
\bvolume{110}
\bpages{9812}.
\end{barticle}
%
%\OrigBibText
%%
%\begin{barticle}[author]
%\bauthor{\bsnm{Ceperley},~\bfnm{DM}\binits{D.}} \AND
%\bauthor{\bsnm{Dewing},~\bfnm{M}\binits{M.}}
%(\byear{1999}).
%\btitle{The penalty method for random walks with uncertain energies}.
%\bjournal{The Journal of chemical physics}
%\bvolume{110}
%\bpages{9812}.
%\end{barticle}
%%
%\endOrigBibText
\bptok{imsref}%
\endbibitem

%b12 ###
\bibitem[\protect\citeauthoryear{Chen, Fox and Guestrin}{2014}]{chen2014stochastic}
%
\begin{binproceedings}[author]
\bauthor{\bsnm{Chen},~\bfnm{Tianqi}\binits{T.}},
\bauthor{\bsnm{Fox},~\bfnm{Emily~B.}\binits{E.~B.}} \AND
\bauthor{\bsnm{Guestrin},~\bfnm{Carlos}\binits{C.}}
(\byear{2014}).
\btitle{Stochastic gradient Hamiltonian {M}onte {C}arlo}.
In \bbooktitle{Proceedings of the 31st International Conference on
Machine Learning (ICML-14)}
\bpages{1683--1691}.
\end{binproceedings}
%
%\OrigBibText
%%
%\begin{binproceedings}[author]
%\bauthor{\bsnm{Chen},~\bfnm{Tianqi}\binits{T.}},
%\bauthor{\bsnm{Fox},~\bfnm{Emily~B}\binits{E.~B.}} \AND
%\bauthor{\bsnm{Guestrin},~\bfnm{Carlos}\binits{C.}}
%(\byear{2014}).
%\btitle{Stochastic Gradient Hamiltonian {M}onte {C}arlo}.
%In \bbooktitle{Proceedings of The 31st International Conference on Machine
%Learning}.
%\end{binproceedings}
%%
%\endOrigBibText
\bptok{imsref}%
\endbibitem

%b13 ###
\bibitem[\protect\citeauthoryear{Del~Moral, Doucet and
Jasra}{2007}]{delmodoucetjasra2007sequential}
%
\begin{bincollection}[mr]
\bauthor{\bsnm{Del Moral},~\bfnm{Pierre}\binits{P.}},
\bauthor{\bsnm{Doucet},~\bfnm{Arnaud}\binits{A.}} \AND
\bauthor{\bsnm{Jasra},~\bfnm{Ajay}\binits{A.}}
(\byear{2007}).
\btitle{Sequential {M}onte {C}arlo for {B}ayesian computation}.
In \bbooktitle{Bayesian Statistics \textbf{8}: Proceedings of the Eighth
Valencia International Meeting, June 2--6, 2006}
(\beditor{\bfnm{J.~M}\binits{J.~M.}~\bsnm{Bernardo}},
\beditor{\bfnm{M.~J.}\binits{M.~J.}~\bsnm{Bayarri}},
\beditor{\bfnm{J.~O.}\binits{J.~O.}~\bsnm{Degroot}},
\beditor{\bfnm{A.~P.}\binits{A.~P.}~\bsnm{Dawid}},
\beditor{\bfnm{D.}\binits{D.}~\bsnm{Heckerman}},
\beditor{\bfnm{A.~M.}\binits{A.~M.}~\bsnm{Smith}} \AND
\beditor{\bfnm{M.}\binits{M.}~\bsnm{West}}, eds.)
\bpages{115--148}.
\bpublisher{Oxford Univ. Press},
\blocation{Oxford}.
\bid{mr={2433191}}
\end{bincollection}
%
%\OrigBibText
%%
%\begin{binproceedings}[author]
%\bauthor{\bsnm{Del~Moral},~\bfnm{Pierre}\binits{P.}},
%\bauthor{\bsnm{Doucet},~\bfnm{Arnaud}\binits{A.}} \AND
%\bauthor{\bsnm{Jasra},~\bfnm{Ajay}\binits{A.}}
%(\byear{2007}).
%\btitle{Sequential {M}onte {C}arlo for bayesian computation}.
%In \bbooktitle{Bayesian statistics 8: proceedings of the eighth Valencia
%International Meeting, June 2-6, 2006}
%(\beditor{\bfnm{J.~M}\binits{J.~M.}~\bsnm{Bernardo}},
%\beditor{\bfnm{M.~J.}\binits{M.~J.}~\bsnm{Bayarri}},
%\beditor{\bfnm{J.~O.}\binits{J.~O.}~\bsnm{Degroot}},
%\beditor{\bfnm{A.~P.}\binits{A.~P.}~\bsnm{Dawid}},
%\beditor{\bfnm{D.}\binits{D.}~\bsnm{Heckerman}},
%\beditor{\bfnm{A.~M.}\binits{A.~M.}~\bsnm{Smith}} \AND
%\beditor{\bfnm{M.}\binits{M.}~\bsnm{West}}, eds.)
%\bvolume{8}
%\bpages{115}.
%\bpublisher{Oxford University Press, USA}.
%\end{binproceedings}
%%
%\endOrigBibText
\bptok{imsref}%
\endbibitem

%b14 ###
\bibitem[\protect\citeauthoryear{Everitt}{2012}]{Everitt2012}
%
\begin{barticle}[mr]
\bauthor{\bsnm{Everitt},~\bfnm{Richard~G.}\binits{R.~G.}}
(\byear{2012}).
\btitle{Bayesian parameter estimation for latent {M}arkov random
fields and social networks}.
\bjournal{J. Comput. Graph. Statist.}
\bvolume{21}
\bpages{940--960}.
\bid{doi={10.1080/10618600.2012.687493}, issn={1061-8600}, mr={3005805}}
\end{barticle}
%
%\OrigBibText
%%
%\begin{barticle}[author]
%\bauthor{\bsnm{Everitt},~\bfnm{Richard~G.}\binits{R.~G.}}
%(\byear{2012}).
%\btitle{Bayesian Parameter Estimation for Latent Markov Random Fields and
%Social Networks}.
%\bjournal{Journal of Computational and Graphical Statistics}
%\bvolume{21}
%\bpages{940-960}.
%\bdoi{10.1080/10618600.2012.687493}
%\end{barticle}
%%
%\endOrigBibText
\bptok{imsref}%
% NOT OUTPUTTED:
% number = 4
% doi = http://dx.doi.org/10.1080/10618600.2012.687493
% fjournal = Journal of Computational and Graphical Statistics
\endbibitem

%b15 ###
\bibitem[\protect\citeauthoryear{Fearnhead, Papaspiliopoulos and
Roberts}{2008}]{fearnhead2008particle}
%
\begin{barticle}[mr]
\bauthor{\bsnm{Fearnhead},~\bfnm{Paul}\binits{P.}},
\bauthor{\bsnm{Papaspiliopoulos},~\bfnm{Omiros}\binits{O.}} \AND
\bauthor{\bsnm{Roberts},~\bfnm{Gareth~O.}\binits{G.~O.}}
(\byear{2008}).
\btitle{Particle filters for partially observed diffusions}.
\bjournal{J. R. Stat. Soc. Ser. B Stat. Methodol.}
\bvolume{70}
\bpages{755--777}.
\bid{doi={10.1111/j.1467-9868.2008.00661.x}, issn={1369-7412}, mr={2523903}}
\end{barticle}
%%
%\OrigBibText
%%
%\begin{barticle}[author]
%\bauthor{\bsnm{Fearnhead},~\bfnm{Paul}\binits{P.}},
%\bauthor{\bsnm{Papaspiliopoulos},~\bfnm{Omiros}\binits{O.}} \AND
%\bauthor{\bsnm{Roberts},~\bfnm{Gareth~O}\binits{G.~O.}}
%(\byear{2008}).
%\btitle{Particle filters for partially observed diffusions}.
%\bjournal{J. Royal Statist. Society Series B}
%\bvolume{70}
%\bpages{755--777}.
%\end{barticle}
%%
%\endOrigBibText
\bptok{imsref}%
% NOT OUTPUTTED:
% number = 4
% doi = http://dx.doi.org/10.1111/j.1467-9868.2008.00661.x
% fjournal = Journal of the Royal Statistical Society. Series B.
%Statistical Methodology
\endbibitem

%b16 ###
\bibitem[\protect\citeauthoryear{Fearnhead et~al.}{2010}]{papasp2010}
%
\begin{barticle}[mr]
\bauthor{\bsnm{Fearnhead},~\bfnm{Paul}\binits{P.}},
\bauthor{\bsnm{Papaspiliopoulos},~\bfnm{Omiros}\binits{O.}},
\bauthor{\bsnm{Roberts},~\bfnm{Gareth~O.}\binits{G.~O.}} \AND
\bauthor{\bsnm{Stuart},~\bfnm{Andrew}\binits{A.}}
(\byear{2010}).
\btitle{Random-weight particle filtering of continuous time processes}.
\bjournal{J. R. Stat. Soc. Ser. B Stat. Methodol.}
\bvolume{72}
\bpages{497--512}.
\bid{doi={10.1111/j.1467-9868.2010.00744.x}, issn={1369-7412}, mr={2758525}}
\end{barticle}
%
%\OrigBibText
%%
%\begin{barticle}[author]
%\bauthor{\bsnm{Fearnhead},~\bfnm{P.}\binits{P.}},
%\bauthor{\bsnm{Papaspiliopoulos},~\bfnm{O.}\binits{O.}},
%\bauthor{\bsnm{Roberts},~\bfnm{G.~O.}\binits{G.~O.}} \AND
%\bauthor{\bsnm{Stuart},~\bfnm{A.}\binits{A.}}
%(\byear{2010}).
%\btitle{Random weight particle filtering of continuous time processes}.
%\bjournal{J. Royal Statist. Society Series B}
%\bvolume{72}
%\bpages{497--513}.
%\end{barticle}
%%
%\endOrigBibText
\bptok{imsref}%
% NOT OUTPUTTED:
% number = 4
% doi = http://dx.doi.org/10.1111/j.1467-9868.2010.00744.x
% fjournal = Journal of the Royal Statistical Society. Series B.
%Statistical Methodology
\endbibitem

%b17 ###
\bibitem[\protect\citeauthoryear{Flegal and Herbei}{2012}]{flegal2012exact}
%
\begin{barticle}[mr]
\bauthor{\bsnm{Flegal},~\bfnm{James~M.}\binits{J.~M.}} \AND
\bauthor{\bsnm{Herbei},~\bfnm{Radu}\binits{R.}}
(\byear{2012}).
\btitle{Exact sampling for intractable probability distributions via a
{B}ernoulli factory}.
\bjournal{Electron. J. Stat.}
\bvolume{6}
\bpages{10--37}.
\bid{doi={10.1214/11-EJS663}, issn={1935-7524}, mr={2879671}}
\end{barticle}
%%
%\OrigBibText
%%
%\begin{barticle}[author]
%\bauthor{\bsnm{Flegal},~\bfnm{James~M}\binits{J.~M.}} \AND
%\bauthor{\bsnm{Herbei},~\bfnm{Radu}\binits{R.}}
%(\byear{2012}).
%\btitle{Exact sampling for intractable probability distributions via a
%Bernoulli factory}.
%\bjournal{Electronic Journal of Statistics}
%\bvolume{6}
%\bpages{10--37}.
%\end{barticle}
%%
%\endOrigBibText
\bptok{imsref}%
% NOT OUTPUTTED:
% doi = http://dx.doi.org/10.1214/11-EJS663
% fjournal = Electronic Journal of Statistics
\endbibitem

%b18 ###
\bibitem[\protect\citeauthoryear{Frei and K{\"
u}nsch}{2013}]{frei2013bridging}
%
\begin{barticle}[mr]
\bauthor{\bsnm{Frei},~\bfnm{M.}\binits{M.}} \AND
\bauthor{\bsnm{K{\"u}nsch},~\bfnm{H.~R.}\binits{H.~R.}}
(\byear{2013}).
\btitle{Bridging the ensemble {K}alman and particle filters}.
\bjournal{Biometrika}
\bvolume{100}
\bpages{781--800}.
\bid{doi={10.1093/biomet/ast020}, issn={0006-3444}, mr={3142332}}
\end{barticle}
%
%\OrigBibText
%%
%\begin{barticle}[author]
%\bauthor{\bsnm{Frei},~\bfnm{Marco}\binits{M.}} \AND
%\bauthor{\bsnm{K{\"u}nsch},~\bfnm{Hans~R}\binits{H.~R.}}
%(\byear{2013}).
%\btitle{Bridging the ensemble Kalman and particle filters}.
%\bjournal{Biometrika}
%\bvolume{100}
%\bpages{781--800}.
%\end{barticle}
%%
%\endOrigBibText
\bptok{imsref}%
% NOT OUTPUTTED:
% number = 4
% doi = http://dx.doi.org/10.1093/biomet/ast020
% fjournal = Biometrika
\endbibitem

%b19 ###
\bibitem[\protect\citeauthoryear{Girolami et~al.}{2013}]{Girolami2013}
%
\begin{bmisc}[author]
\bauthor{\bsnm{Girolami},~\bfnm{M.}\binits{M.}},
\bauthor{\bsnm{Lyne},~\bfnm{A.~M.}\binits{A.~M.}},
\bauthor{\bsnm{Strathmann},~\bfnm{H.}\binits{H.}},
\bauthor{\bsnm{Simpson},~\bfnm{D.}\binits{D.}} \AND
\bauthor{\bsnm{Atchade},~\bfnm{Y.}\binits{Y.}}
(\byear{2013}).
\bhowpublished{Playing Russian roulette with intractable likelihoods.
Preprint. Available at \arxivurl{arXiv:1306.4032}.}
\end{bmisc}
%%
%\OrigBibText
%%
%\begin{barticle}[author]
%\bauthor{\bsnm{{Girolami}},~\bfnm{M.}\binits{M.}},
%\bauthor{\bsnm{{Lyne}},~\bfnm{A.~M.}\binits{A.~M.}},
%\bauthor{\bsnm{{Strathmann}},~\bfnm{H.}\binits{H.}},
%\bauthor{\bsnm{{Simpson}},~\bfnm{D.}\binits{D.}} \AND
%\bauthor{\bsnm{{Atchade}},~\bfnm{Y.}\binits{Y.}}
%(\byear{2013}).
%\btitle{{Playing Russian Roulette with Intractable Likelihoods}}.
%\bjournal{ArXiv e-prints}.
%\end{barticle}
%%
%\endOrigBibText
\bptok{imsref}%
\endbibitem

%b20 ###
\bibitem[\protect\citeauthoryear{Jourdain and Sbai}{2007}]{jourdain2007exact}
%
\begin{barticle}[mr]
\bauthor{\bsnm{Jourdain},~\bfnm{Benjamin}\binits{B.}} \AND
\bauthor{\bsnm{Sbai},~\bfnm{Mohamed}\binits{M.}}
(\byear{2007}).
\btitle{Exact retrospective {M}onte {C}arlo computation of arithmetic
average {A}sian options}.
\bjournal{Monte Carlo Methods Appl.}
\bvolume{13}
\bpages{135--171}.
\bid{doi={10.1515/mcma.2007.008}, issn={0929-9629}, mr={2338086}}
\end{barticle}
%
%\OrigBibText
%%
%\begin{barticle}[author]
%\bauthor{\bsnm{Jourdain},~\bfnm{Benjamin}\binits{B.}} \AND
%\bauthor{\bsnm{Sbai},~\bfnm{Mohamed}\binits{M.}}
%(\byear{2007}).
%\btitle{Exact retrospective {M}onte {C}arlo computation of arithmetic average
%Asian options}.
%\bjournal{{M}onte {C}arlo Methods and Applications}
%\bvolume{13}
%\bpages{135--171}.
%\end{barticle}
%%
%\endOrigBibText
\bptok{imsref}%
% NOT OUTPUTTED:
% number = 2
% doi = http://dx.doi.org/10.1515/mcma.2007.008
% fjournal = Monte Carlo Methods and Applications
\endbibitem

%b21 ###
\bibitem[\protect\citeauthoryear{Keane and O'Brien}{1994}]{Keane1994}
%
\begin{barticle}[author]
\bauthor{\bsnm{Keane},~\bfnm{M.~S.}\binits{M.~S.}} \AND
\bauthor{\bsnm{O'Brien},~\bfnm{George~L.}\binits{G.~L.}}
(\byear{1994}).
\btitle{A Bernoulli factory}.
\bjournal{ACM Trans. Model. Comput. Simul.}
\bvolume{4}
\bpages{213--219}.
\bid{doi={10.1145/175007.175019}}
\end{barticle}
%
%\OrigBibText
%%
%\begin{barticle}[author]
%\bauthor{\bsnm{Keane},~\bfnm{M.~S.}\binits{M.~S.}} \AND
%\bauthor{\bsnm{O'Brien},~\bfnm{George~L.}\binits{G.~L.}}
%(\byear{1994}).
%\btitle{A Bernoulli factory}.
%\bjournal{ACM Trans. Model. Comput. Simul.}
%\bvolume{4}
%\bpages{213--219}.
%\bdoi{10.1145/175007.175019}
%\end{barticle}
%%
%\endOrigBibText
\bptok{imsref}%
\endbibitem

%b22 ###
\bibitem[\protect\citeauthoryear{Kennedy and Kuti}{1985}]{kennedykuti1985}
%
\begin{barticle}[pbm]
\bauthor{\bsnm{Kennedy},~\bfnm{A.~D.}\binits{A.~D.}} \AND
\bauthor{\bsnm{Kuti},~\bfnm{J.}\binits{J.}}
(\byear{1985}).
\btitle{Noise without noise: A new Monte Carlo method}.
\bjournal{Phys. Rev. Lett.}
\bvolume{54}
\bpages{2473--2476}.
\bid{issn={1079-7114}, pmid={10031352}}
\end{barticle}
%
%\OrigBibText
%%
%\begin{barticle}[author]
%\bauthor{\bsnm{Kennedy},~\bfnm{A.~D.}\binits{A.~D.}} \AND
%\bauthor{\bsnm{Kuti},~\bfnm{J.}\binits{J.}}
%(\byear{1985}).
%\btitle{Noise without Noise: A New {M}onte {C}arlo Method}.
%\bjournal{Phys. Rev. Lett.}
%\bvolume{54}
%\bpages{2473--2476}.
%\bdoi{10.1103/PhysRevLett.54.2473}
%\end{barticle}
%%
%\endOrigBibText
\bptok{imsref}%
% NOT OUTPUTTED:
% number = 23
% fjournal = Physical review letters
\endbibitem

%b23 ###
\bibitem[\protect\citeauthoryear{Kleiner et~al.}{2014}]{kleiner2011scalable}
%
\begin{barticle}[author]
\bauthor{\bsnm{Kleiner},~\bfnm{Ariel}\binits{A.}},
\bauthor{\bsnm{Talwalkar},~\bfnm{Ameet}\binits{A.}},
\bauthor{\bsnm{Sarkar},~\bfnm{Purnamrita}\binits{P.}} \AND
\bauthor{\bsnm{Jordan},~\bfnm{Michael~I.}\binits{M.~I.}}
(\byear{2014}).
\btitle{A scalable bootstrap for massive data}.
\bjournal{J. Roy. Statist. Soc. Ser. B}
\bvolume{76}
\bpages{795--816}.
\end{barticle}
%%
%\OrigBibText
%%
%\begin{barticle}[author]
%\bauthor{\bsnm{Kleiner},~\bfnm{Ariel}\binits{A.}},
%\bauthor{\bsnm{Talwalkar},~\bfnm{Ameet}\binits{A.}},
%\bauthor{\bsnm{Sarkar},~\bfnm{Purnamrita}\binits{P.}} \AND
%\bauthor{\bsnm{Jordan},~\bfnm{Michael~I.}\binits{M.~I.}}
%(\byear{2014}).
%\btitle{A scalable bootstrap for massive data}.
%\bjournal{J. Royal Statist. Society Series B}.
%\bdoi{10.1111/rssb.12050}
%\end{barticle}
%%
%\endOrigBibText
\bptok{imsref}%
\endbibitem

%b24 ###
\bibitem[\protect\citeauthoryear{Kuti}{1982}]{kuti1982stochastic}
%
\begin{barticle}[author]
\bauthor{\bsnm{Kuti},~\bfnm{Julius}\binits{J.}}
(\byear{1982}).
\btitle{Stochastic method for the numerical study of lattice fermions}.
\bjournal{Phys. Rev. Lett.}
\bvolume{49}
\bpages{183--186}.
\end{barticle}
%%
%\OrigBibText
%%
%\begin{barticle}[author]
%\bauthor{\bsnm{Kuti},~\bfnm{Julius}\binits{J.}}
%(\byear{1982}).
%\btitle{Stochastic method for the numerical study of lattice fermions}.
%\bjournal{Physical Review Letters}
%\bvolume{49}
%\bpages{183}.
%\end{barticle}
%%
%\endOrigBibText
\bptok{imsref}%
\endbibitem

%b25 ###
\bibitem[\protect\citeauthoryear{{\L}atuszy{\'n}ski
et~al.}{2011}]{latuszynski2011simulating}
%
\begin{barticle}[mr]
\bauthor{\bsnm{{\L}atuszy{\'n}ski},~\bfnm{Krzysztof}\binits{K.}},
\bauthor{\bsnm{Kosmidis},~\bfnm{Ioannis}\binits{I.}},
\bauthor{\bsnm{Papaspiliopoulos},~\bfnm{Omiros}\binits{O.}} \AND
\bauthor{\bsnm{Roberts},~\bfnm{Gareth~O.}\binits{G.~O.}}
(\byear{2011}).
\btitle{Simulating events of unknown probabilities via reverse time
martingales}.
\bjournal{Random Structures Algorithms}
\bvolume{38}
\bpages{441--452}.
\bid{doi={10.1002/rsa.20333}, issn={1042-9832}, mr={2829311}}
\end{barticle}
%
%\OrigBibText
%%
%\begin{barticle}[author]
%\bauthor{\bsnm{{\L}atuszy{\'n}ski},~\bfnm{Krzysztof}\binits{K.}},
%\bauthor{\bsnm{Kosmidis},~\bfnm{Ioannis}\binits{I.}},
%\bauthor{\bsnm{Papaspiliopoulos},~\bfnm{Omiros}\binits{O.}} \AND
%\bauthor{\bsnm{Roberts},~\bfnm{Gareth~O}\binits{G.~O.}}
%(\byear{2011}).
%\btitle{Simulating events of unknown probabilities via reverse time
%martingales}.
%\bjournal{Random Structures \& Algorithms}
%\bvolume{38}
%\bpages{441--452}.
%\end{barticle}
%%
%\endOrigBibText
\bptok{imsref}%
% NOT OUTPUTTED:
% number = 4
% doi = http://dx.doi.org/10.1002/rsa.20333
% fjournal = Random Structures \& Algorithms
\endbibitem

%b26 ###
\bibitem[\protect\citeauthoryear{Lin, Liu and Sloan}{2000}]{lin2000noisy}
%
\begin{barticle}[author]
\bauthor{\bsnm{Lin},~\bfnm{L.}\binits{L.}},
\bauthor{\bsnm{Liu},~\bfnm{K.~F.}\binits{K.~F.}} \AND
\bauthor{\bsnm{Sloan},~\bfnm{J.}\binits{J.}}
(\byear{2000}).
\btitle{A noisy {M}onte {C}arlo algorithm}.
\bjournal{Phys. Rev. D}
\bvolume{61}
\bpages{074505}.
\end{barticle}
%
%\OrigBibText
%%
%\begin{barticle}[author]
%\bauthor{\bsnm{Lin},~\bfnm{L}\binits{L.}},
%\bauthor{\bsnm{Liu},~\bfnm{KF}\binits{K.}} \AND
%\bauthor{\bsnm{Sloan},~\bfnm{J}\binits{J.}}
%(\byear{2000}).
%\btitle{A noisy {M}onte {C}arlo algorithm}.
%\bjournal{Physical Review D}
%\bvolume{61}
%\bpages{074505}.
%\end{barticle}
%%
%\endOrigBibText
\bptok{imsref}%
\endbibitem

%b27 ###
\bibitem[\protect\citeauthoryear{Liu and Chen}{1998}]{liu1998sequential}
%
\begin{barticle}[mr]
\bauthor{\bsnm{Liu},~\bfnm{Jun~S.}\binits{J.~S.}} \AND
\bauthor{\bsnm{Chen},~\bfnm{Rong}\binits{R.}}
(\byear{1998}).
\btitle{Sequential {M}onte {C}arlo methods for dynamic systems}.
\bjournal{J. Amer. Statist. Assoc.}
\bvolume{93}
\bpages{1032--1044}.
\bid{doi={10.2307/2669847}, issn={0162-1459}, mr={1649198}}
\end{barticle}
%
%\OrigBibText
%%
%\begin{barticle}[author]
%\bauthor{\bsnm{Liu},~\bfnm{Jun~S}\binits{J.~S.}} \AND
%\bauthor{\bsnm{Chen},~\bfnm{Rong}\binits{R.}}
%(\byear{1998}).
%\btitle{Sequential {M}onte {C}arlo methods for dynamic systems}.
%\bjournal{Journal of the American statistical association}
%\bvolume{93}
%\bpages{1032--1044}.
%\end{barticle}
%%
%\endOrigBibText
\bptok{imsref}%
% NOT OUTPUTTED:
% number = 443
% doi = http://dx.doi.org/10.2307/2669847
% coden = JSTNAL
% fjournal = Journal of the American Statistical Association
\endbibitem

%b28 ###
\bibitem[\protect\citeauthoryear{Maclaurin and
Adams}{2014}]{maclaurin2014firefly}
%
\begin{bmisc}[author]
\bauthor{\bsnm{Maclaurin},~\bfnm{Dougal}\binits{D.}} \AND
\bauthor{\bsnm{Adams},~\bfnm{Ryan~P.}\binits{R.~P.}}
(\byear{2014}).
\bhowpublished{Firefly {M}onte {C}arlo: Exact MCMC with subsets of data.
Preprint. Available at \arxivurl{arXiv:1403.5693}.}
\end{bmisc}
%
%\OrigBibText
%%
%\begin{barticle}[author]
%\bauthor{\bsnm{Maclaurin},~\bfnm{Dougal}\binits{D.}} \AND
%\bauthor{\bsnm{Adams},~\bfnm{Ryan~P}\binits{R.~P.}}
%(\byear{2014}).
%\btitle{Firefly {M}onte {C}arlo: Exact MCMC with Subsets of Data}.
%\bjournal{arXiv preprint arXiv:1403.5693}.
%\end{barticle}
%%
%\endOrigBibText
\bptok{imsref}%
\endbibitem

%b29 ###
\bibitem[\protect\citeauthoryear{Marin et~al.}{2012}]{marin2012approximate}
%
\begin{barticle}[mr]
\bauthor{\bsnm{Marin},~\bfnm{Jean-Michel}\binits{J.-M.}},
\bauthor{\bsnm{Pudlo},~\bfnm{Pierre}\binits{P.}},
\bauthor{\bsnm{Robert},~\bfnm{Christian~P.}\binits{C.~P.}} \AND
\bauthor{\bsnm{Ryder},~\bfnm{Robin~J.}\binits{R.~J.}}
(\byear{2012}).
\btitle{Approximate {B}ayesian computational methods}.
\bjournal{Stat. Comput.}
\bvolume{22}
\bpages{1167--1180}.
\bid{doi={10.1007/s11222-011-9288-2}, issn={0960-3174}, mr={2992292}}
\end{barticle}
%
%\OrigBibText
%%
%\begin{barticle}[author]
%\bauthor{\bsnm{Marin},~\bfnm{Jean-Michel}\binits{J.-M.}},
%\bauthor{\bsnm{Pudlo},~\bfnm{Pierre}\binits{P.}},
%\bauthor{\bsnm{Robert},~\bfnm{Christian~P}\binits{C.~P.}} \AND
%\bauthor{\bsnm{Ryder},~\bfnm{Robin~J}\binits{R.~J.}}
%(\byear{2012}).
%\btitle{Approximate Bayesian computational methods}.
%\bjournal{Statistics and Computing}
%\bvolume{22}
%\bpages{1167--1180}.
%\end{barticle}
%%
%\endOrigBibText
\bptok{imsref}%
% NOT OUTPUTTED:
% number = 6
% doi = http://dx.doi.org/10.1007/s11222-011-9288-2
% coden = STACE3
% fjournal = Statistics and Computing
\endbibitem

%b30 ###
\bibitem[\protect\citeauthoryear{McLeish}{2011}]{McLeish2011}
%
\begin{barticle}[mr]
\bauthor{\bsnm{McLeish},~\bfnm{Don}\binits{D.}}
(\byear{2011}).
\btitle{A general method for debiasing a {M}onte {C}arlo estimator}.
\bjournal{Monte Carlo Methods Appl.}
\bvolume{17}
\bpages{301--315}.
\bid{doi={10.1515/mcma.2011.013}, issn={0929-9629}, mr={2890424}}
\end{barticle}
%
%\OrigBibText
%%
%\begin{barticle}[author]
%\bauthor{\bsnm{{McLeish}},~\bfnm{D.}\binits{D.}}
%(\byear{2011}).
%\btitle{{A general method for debiasing a {M}onte {C}arlo estimator}}.
%\bjournal{{M}onte {C}arlo Methods and Applications}
%\bvolume{17}
%\bpages{301--315}.
%\end{barticle}
%%
%\endOrigBibText
\bptok{imsref}%
% NOT OUTPUTTED:
% number = 4
% doi = http://dx.doi.org/10.1515/mcma.2011.013
% fjournal = Monte Carlo Methods and Applications
\endbibitem

%b31 ###
\bibitem[\protect\citeauthoryear{M{\o}ller
et~al.}{2006}]{moller2006efficient}
%
\begin{barticle}[mr]
\bauthor{\bsnm{M{\o}ller},~\bfnm{J.}\binits{J.}},
\bauthor{\bsnm{Pettitt},~\bfnm{A.~N.}\binits{A.~N.}},
\bauthor{\bsnm{Reeves},~\bfnm{R.}\binits{R.}} \AND
\bauthor{\bsnm{Berthelsen},~\bfnm{K.~K.}\binits{K.~K.}}
(\byear{2006}).
\btitle{An efficient {M}arkov chain {M}onte {C}arlo method for
distributions with intractable normalising constants}.
\bjournal{Biometrika}
\bvolume{93}
\bpages{451--458}.
\bid{doi={10.1093/biomet/93.2.451}, issn={0006-3444}, mr={2278096}}
\end{barticle}
%
%\OrigBibText
%%
%\begin{barticle}[author]
%\bauthor{\bsnm{M{\o}ller},~\bfnm{Jesper}\binits{J.}},
%\bauthor{\bsnm{Pettitt},~\bfnm{Anthony~N}\binits{A.~N.}},
%\bauthor{\bsnm{Reeves},~\bfnm{R}\binits{R.}} \AND
%\bauthor{\bsnm{Berthelsen},~\bfnm{Kasper~Klitgaard}\binits{K.~K.}}
%(\byear{2006}).
%\btitle{An efficient Markov chain {M}onte {C}arlo method for
%distributions with
%intractable normalising constants}.
%\bjournal{Biometrika}
%\bvolume{93}
%\bpages{451--458}.
%\end{barticle}
%%
%\endOrigBibText
\bptok{imsref}%
% NOT OUTPUTTED:
% number = 2
% doi = http://dx.doi.org/10.1093/biomet/93.2.451
% coden = BIOKAX
% fjournal = Biometrika
\endbibitem

%b32 ###
\bibitem[\protect\citeauthoryear{Nacu and Peres}{2005}]{Nacu2005}
%
\begin{barticle}[mr]
\bauthor{\bsnm{Nacu},~\bfnm{{\c{S}}erban}\binits{\c{S}.}} \AND
\bauthor{\bsnm{Peres},~\bfnm{Yuval}\binits{Y.}}
(\byear{2005}).
\btitle{Fast simulation of new coins from old}.
\bjournal{Ann. Appl. Probab.}
\bvolume{15}
\bpages{93--115}.
\bid{doi={10.1214/105051604000000549}, issn={1050-5164}, mr={2115037}}
\end{barticle}
%
%\OrigBibText
%%
%\begin{barticle}[author]
%\bauthor{\bsnm{Nacu},~\bfnm{{\c{S}}erban}\binits{{\c{S}}.}} \AND
%\bauthor{\bsnm{Peres},~\bfnm{Yuval}\binits{Y.}}
%(\byear{2005}).
%\btitle{Fast simulation of new coins from old}.
%\bjournal{Ann. Applied Probability}
%\bvolume{15}
%\bpages{93--115}.
%\end{barticle}
%%
%\endOrigBibText
\bptok{imsref}%
% NOT OUTPUTTED:
% number = 1A
% doi = http://dx.doi.org/10.1214/105051604000000549
% fjournal = The Annals of Applied Probability
\endbibitem

%b33 ###
\bibitem[\protect\citeauthoryear{Nicholls, Fox and Watt}{2012}]{nicholls2012coupled}
%
\begin{bmisc}[author]
\bauthor{\bsnm{Nicholls},~\bfnm{Geoff~K.}\binits{G.~K.}},
\bauthor{\bsnm{Fox},~\bfnm{Colin}\binits{C.}} \AND
\bauthor{\bsnm{Watt},~\bfnm{Alexis~Muir}\binits{A.~M.}}
(\byear{2012}).
\bhowpublished{Coupled MCMC with a randomized acceptance probability.
Preprint. Available at \arxivurl{arXiv:1205.6857}.}
\end{bmisc}
%%
%\OrigBibText
%%
%\begin{barticle}[author]
%\bauthor{\bsnm{Nicholls},~\bfnm{Geoff~K}\binits{G.~K.}},
%\bauthor{\bsnm{Fox},~\bfnm{Colin}\binits{C.}} \AND
%\bauthor{\bsnm{Watt},~\bfnm{Alexis~Muir}\binits{A.~M.}}
%(\byear{2012}).
%\btitle{Coupled MCMC with a randomized acceptance probability}.
%\bjournal{ArXiv e-prints}.
%\end{barticle}
%%
%\endOrigBibText
\bptok{imsref}%
\endbibitem

%b34 ###
\bibitem[\protect\citeauthoryear{Olsson and Str{\"
o}jby}{2011}]{olsson2011particle}
%
\begin{barticle}[mr]
\bauthor{\bsnm{Olsson},~\bfnm{Jimmy}\binits{J.}} \AND
\bauthor{\bsnm{Str{\"o}jby},~\bfnm{Jonas}\binits{J.}}
(\byear{2011}).
\btitle{Particle-based likelihood inference in partially observed
diffusion processes using generalised {P}oisson estimators}.
\bjournal{Electron. J. Stat.}
\bvolume{5}
\bpages{1090--1122}.
\bid{doi={10.1214/11-EJS632}, issn={1935-7524}, mr={2836770}}
\end{barticle}
%
%\OrigBibText
%%
%\begin{barticle}[author]
%\bauthor{\bsnm{Olsson},~\bfnm{Jimmy}\binits{J.}} \AND
%\bauthor{\bsnm{Str{\"o}jby},~\bfnm{Jonas}\binits{J.}}
%(\byear{2011}).
%\btitle{Particle-based likelihood inference in partially observed diffusion
%processes using generalised Poisson estimators}.
%\bjournal{Electronic Journal of Statistics}
%\bvolume{5}
%\bpages{1090--1122}.
%\end{barticle}
%%
%\endOrigBibText
\bptok{imsref}%
% NOT OUTPUTTED:
% doi = http://dx.doi.org/10.1214/11-EJS632
% fjournal = Electronic Journal of Statistics
\endbibitem

%b35 ###
\bibitem[\protect\citeauthoryear{Papaspiliopoulos}{2011}]{papasp2011bayesian}
%
\begin{binproceedings}[mr]
\bauthor{\bsnm{Papaspiliopoulos},~\bfnm{O.}\binits{O.}}
(\byear{2011}).
\btitle{A methodological framework for {M}onte {C}arlo probabilistic inference
for diffusion processes}.
In \bbooktitle{Bayesian Time Series Models}
(\beditor{\bfnm{David}\binits{D.}~\bsnm{Barber}},
\beditor{\bfnm{A~Taylan}\binits{A.~T.}~\bsnm{Cemgil}} \AND
\beditor{\bfnm{Silvia}\binits{S.}~\bsnm{Chiappa}}, eds.)
\bpages{82--99}.
\bpublisher{Cambridge Univ. Press},
\blocation{Cambridge}.
\bid{doi={10.1017/CBO9780511984679}, mr={2894230}}
\end{binproceedings}
%
%\OrigBibText
%%
%\begin{binproceedings}[author]
%\bauthor{\bsnm{Papaspiliopoulos},~\bfnm{O.}\binits{O.}}
%(\byear{2011}).
%\btitle{A methodological framework for {M}onte {C}arlo probabilistic inference
%for diffusion processes}.
%In \bbooktitle{Bayesian time series models}
%(\beditor{\bfnm{David}\binits{D.}~\bsnm{Barber}},
%\beditor{\bfnm{A~Taylan}\binits{A.~T.}~\bsnm{Cemgil}} \AND
%\beditor{\bfnm{Silvia}\binits{S.}~\bsnm{Chiappa}}, eds.).
%\bpublisher{Cambridge University Press}.
%\end{binproceedings}
%%
%\endOrigBibText
\bptok{imsref}%
% NOT OUTPUTTED:
% doi = http://dx.doi.org/10.1017/CBO9780511984679
% isbn = 978-0-521-19676-5
% fpage = xiv+417
\endbibitem

%b36 ###
\bibitem[\protect\citeauthoryear{Rhee and Glynn}{2012}]{RheeGlynn2012}
%
\begin{binproceedings}[author]
\bauthor{\bsnm{Rhee},~\bfnm{Chang-han}\binits{C.-h.}} \AND
\bauthor{\bsnm{Glynn},~\bfnm{Peter~W.}\binits{P.~W.}}
(\byear{2012}).
\btitle{A new approach to unbiased estimation for SDE's}.
In \bbooktitle{Proceedings of the Winter Simulation Conference}.
\bpages{17:1--17:7}.
\bpublisher{Winter Simulation Conference},
\blocation{Berlin}.
\end{binproceedings}
%
%\OrigBibText
%%
%\begin{binproceedings}[author]
%\bauthor{\bsnm{Rhee},~\bfnm{Chang-han}\binits{C.-h.}} \AND
%\bauthor{\bsnm{Glynn},~\bfnm{Peter~W.}\binits{P.~W.}}
%(\byear{2012}).
%\btitle{A new approach to unbiased estimation for SDE's}.
%In \bbooktitle{Proceedings of the Winter Simulation Conference}.
%\bseries{WSC '12}
%\bpages{17:1--17:7}.
%\bpublisher{Winter Simulation Conference}.
%\end{binproceedings}
%%
%\endOrigBibText
\bptok{imsref}%
\endbibitem

%b37 ###
\bibitem[\protect\citeauthoryear{Rhee and Glynn}{2013}]{RheeGlynn2013}
%
\begin{bmisc}[author]
\bauthor{\bsnm{Rhee},~\bfnm{Chang-han}\binits{C.-h.}} \AND
\bauthor{\bsnm{Glynn},~\bfnm{Peter~W.}\binits{P.~W.}}
(\byear{2013}).
\bhowpublished{Unbiased estimation with square root convergence for
S{D}{E} models.
Technical report, Stanford Univ., Stanford, CA.}
\end{bmisc}
%
%\OrigBibText
%%
%\begin{btechreport}[author]
%\bauthor{\bsnm{Rhee},~\bfnm{Chang-han}\binits{C.-h.}} \AND
%\bauthor{\bsnm{Glynn},~\bfnm{Peter~W.}\binits{P.~W.}}
%(\byear{2013}).
%\btitle{{U}nbiased {E}stimation with {S}quare {R}oot {C}onvergence for
%{S}{D}{E} {M}odels.}
%\btype{Technical Report},
%\bpublisher{Stanford University}.
%\end{btechreport}
%%
%\endOrigBibText
\bptok{imsref}%
% NOT OUTPUTTED:
% publisher = Stanford Univ.
\endbibitem

%b38 ###
\bibitem[\protect\citeauthoryear{Rychlik}{1990}]{rychlik1990unbiased}
%
\begin{barticle}[mr]
\bauthor{\bsnm{Rychlik},~\bfnm{Tomasz}\binits{T.}}
(\byear{1990}).
\btitle{Unbiased nonparametric estimation of the derivative of the mean}.
\bjournal{Statist. Probab. Lett.}
\bvolume{10}
\bpages{329--333}.
\bid{doi={10.1016/0167-7152(90)90051-8}, issn={0167-7152}, mr={1069911}}
\end{barticle}
%
%\OrigBibText
%%
%\begin{barticle}[author]
%\bauthor{\bsnm{Rychlik},~\bfnm{Tomasz}\binits{T.}}
%(\byear{1990}).
%\btitle{Unbiased nonparametric estimation of the derivative of the mean}.
%\bjournal{Statistics \& probability letters}
%\bvolume{10}
%\bpages{329--333}.
%\end{barticle}
%%
%\endOrigBibText
\bptok{imsref}%
% NOT OUTPUTTED:
% number = 4
% doi = http://dx.doi.org/10.1016/0167-7152(90)90051-8
% coden = SPLTDC
% fjournal = Statistics \& Probability Letters
\endbibitem

%b39 ###
\bibitem[\protect\citeauthoryear{Rychlik}{1995}]{rychlik1995class}
%
\begin{barticle}[mr]
\bauthor{\bsnm{Rychlik},~\bfnm{T.}\binits{T.}}
(\byear{1995}).
\btitle{A class of unbiased kernel estimates of a probability density
function}.
\bjournal{Appl. Math. (Warsaw)}
\bvolume{22}
\bpages{485--497}.
\bid{issn={1233-7234}, mr={1314912}}
\end{barticle}
%
%\OrigBibText
%%
%\begin{barticle}[author]
%\bauthor{\bsnm{Rychlik},~\bfnm{Tomasz}\binits{T.}}
%(\byear{1995}).
%\btitle{A class of unbiased kernel estimates of a probability density
%function}.
%\bjournal{Applicationes Math}
%\bvolume{22}
%\bpages{485--497}.
%\end{barticle}
%%
%\endOrigBibText
\bptok{imsref}%
% NOT OUTPUTTED:
% number = 4
% fjournal = Applicationes Mathematicae
\endbibitem

%b40 ###
\bibitem[\protect\citeauthoryear{Sermaidis et~al.}{2015}]{sermaidis2012markov}
%
\begin{barticle}[author]
\bauthor{\bsnm{Sermaidis},~\bfnm{Giorgos}\binits{G.}},
\bauthor{\bsnm{Papaspiliopoulos},~\bfnm{Omiros}\binits{O.}},
\bauthor{\bsnm{Roberts},~\bfnm{Gareth~O.}\binits{G.~O.}},
\bauthor{\bsnm{Beskos},~\bfnm{Alexandros}\binits{A.}} \AND
\bauthor{\bsnm{Fearnhead},~\bfnm{Paul}\binits{P.}}
(\byear{2015}).
\btitle{Markov chain {M}onte {C}arlo for exact inference for diffusions}.
\bjournal{Scand. J. Stat.}
\bvolume{40}
\bpages{294--321}.
\end{barticle}
%%
%\OrigBibText
%%
%\begin{barticle}[author]
%\bauthor{\bsnm{Sermaidis},~\bfnm{Giorgos}\binits{G.}},
%\bauthor{\bsnm{Papaspiliopoulos},~\bfnm{Omiros}\binits{O.}},
%\bauthor{\bsnm{Roberts},~\bfnm{Gareth~O}\binits{G.~O.}},
%\bauthor{\bsnm{Beskos},~\bfnm{Alexandros}\binits{A.}} \AND
%\bauthor{\bsnm{Fearnhead},~\bfnm{Paul}\binits{P.}}
%(\byear{2012}).
%\btitle{Markov chain {M}onte {C}arlo for exact inference for diffusions}.
%\bjournal{Scandinavian Journal of Statistics}.
%\end{barticle}
%%
%\endOrigBibText
\bptok{imsref}%
\endbibitem

%b41 ###
\bibitem[\protect\citeauthoryear{Thomas and Blanchet}{2011}]{thomas2011practical}
%
\begin{bmisc}[author]
\bauthor{\bsnm{Thomas},~\bfnm{A.~C.}\binits{A.~C.}} \AND
\bauthor{\bsnm{Blanchet},~\bfnm{Jose~H.}\binits{J.~H.}}
(\byear{2011}).
\bhowpublished{A practical implementation of the Bernoulli factory.
Preprint. Available at \arxivurl{arXiv:1106.2508}.}
\end{bmisc}
%%
%\OrigBibText
%%
%\begin{barticle}[author]
%\bauthor{\bsnm{Thomas},~\bfnm{AC}\binits{A.}} \AND
%\bauthor{\bsnm{Blanchet},~\bfnm{Jose~H}\binits{J.~H.}}
%(\byear{2011}).
%\btitle{A Practical Implementation of the Bernoulli Factory}.
%\bjournal{ArXiv e-prints}.
%\end{barticle}
%%
%\endOrigBibText
\bptok{imsref}%
\endbibitem

%b42 ###
\bibitem[\protect\citeauthoryear{Tran et~al.}{2013}]{tranissquare2013}
%
\begin{bmisc}[author]
\bauthor{\bsnm{Tran},~\bfnm{M.~N.}\binits{M.~N.}},
\bauthor{\bsnm{Scharth},~\bfnm{M.}\binits{M.}},
\bauthor{\bsnm{Pitt},~\bfnm{M.~K.}\binits{M.~K.}} \AND
\bauthor{\bsnm{Kohn},~\bfnm{R.}\binits{R.}}
(\byear{2013}).
\bhowpublished{Importance sampling squared for Bayesian inference in latent variable models
Preprint. Available at \arxivurl{arXiv:1309.3339}.}
\end{bmisc}
%
%\OrigBibText
%%
%\begin{barticle}[author]
%\bauthor{\bsnm{{Tran}},~\bfnm{M.~N.}\binits{M.~N.}},
%\bauthor{\bsnm{{Scharth}},~\bfnm{M.}\binits{M.}},
%\bauthor{\bsnm{{Pitt}},~\bfnm{M.~K.}\binits{M.~K.}} \AND
%\bauthor{\bsnm{{Kohn}},~\bfnm{R.}\binits{R.}}
%(\byear{2013}).
%\btitle{{IS\^{}2 for Bayesian inference in latent variable models}}.
%\bjournal{ArXiv e-prints}.
%\end{barticle}
%%
%\endOrigBibText
\bptok{imsref}%
\endbibitem

%b43 ###
\bibitem[\protect\citeauthoryear{Troyer and
Wiese}{2005}]{troyer2005computational}
%
\begin{barticle}[pbm]
\bauthor{\bsnm{Troyer},~\bfnm{Matthias}\binits{M.}} \AND
\bauthor{\bsnm{Wiese},~\bfnm{Uwe-Jens}\binits{U.-J.}}
(\byear{2005}).
\btitle{Computational complexity and fundamental limitations to
fermionic quantum Monte Carlo simulations}.
\bjournal{Phys. Rev. Lett.}
\bvolume{94}
\bpages{170201}.
\bid{issn={0031-9007}, pmid={15904269}}
\end{barticle}
%
%\OrigBibText
%%
%\begin{barticle}[author]
%\bauthor{\bsnm{Troyer},~\bfnm{Matthias}\binits{M.}} \AND
%\bauthor{\bsnm{Wiese},~\bfnm{Uwe-Jens}\binits{U.-J.}}
%(\byear{2005}).
%\btitle{Computational complexity and fundamental limitations to fermionic
%quantum {M}onte {C}arlo simulations}.
%\bjournal{Physical review letters}
%\bvolume{94}
%\bpages{170201}.
%\end{barticle}
%%
%\endOrigBibText
\bptok{imsref}%
% NOT OUTPUTTED:
% number = 17
% fjournal = Physical review letters
\endbibitem

%b44 ###
\bibitem[\protect\citeauthoryear{Wagner}{1987}]{wagner1987unbiased}
%
\begin{barticle}[mr]
\bauthor{\bsnm{Wagner},~\bfnm{Wolfgang}\binits{W.}}
(\byear{1987}).
\btitle{Unbiased {M}onte {C}arlo evaluation of certain functional integrals}.
\bjournal{J. Comput. Phys.}
\bvolume{71}
\bpages{21--33}.
\bid{doi={10.1016/0021-9991(87)90017-9}, issn={0021-9991}, mr={0895521}}
\end{barticle}
%
%\OrigBibText
%%
%\begin{barticle}[author]
%\bauthor{\bsnm{Wagner},~\bfnm{Wolfgang}\binits{W.}}
%(\byear{1987}).
%\btitle{Unbiased {M}onte {C}arlo evaluation of certain functional integrals}.
%\bjournal{Journal of Computational Physics}
%\bvolume{71}
%\bpages{21--33}.
%\end{barticle}
%%
%\endOrigBibText
\bptok{imsref}%
% NOT OUTPUTTED:
% number = 1
% doi = http://dx.doi.org/10.1016/0021-9991(87)90017-9
% coden = JCTPAH
% fjournal = Journal of Computational Physics
\endbibitem

%b45 ###
\bibitem[\protect\citeauthoryear{Walker}{2011}]{walker2011posterior}
%
\begin{barticle}[mr]
\bauthor{\bsnm{Walker},~\bfnm{Stephen~G.}\binits{S.~G.}}
(\byear{2011}).
\btitle{Posterior sampling when the normalizing constant is unknown}.
\bjournal{Comm. Statist. Simulation Comput.}
\bvolume{40}
\bpages{784--792}.
\bid{doi={10.1080/03610918.2011.555042}, issn={0361-0918}, mr={2783887}}
\end{barticle}
%
%\OrigBibText
%%
%\begin{barticle}[author]
%\bauthor{\bsnm{Walker},~\bfnm{Stephen~G}\binits{S.~G.}}
%(\byear{2011}).
%\btitle{Posterior sampling when the normalizing constant is unknown}.
%\bjournal{Communications in Statistics-Simulation and Computation}
%\bvolume{40}
%\bpages{784--792}.
%\end{barticle}
%%
%\endOrigBibText
\bptok{imsref}%
% NOT OUTPUTTED:
% number = 5
% doi = http://dx.doi.org/10.1080/03610918.2011.555042
% coden = CSSCDB
% fjournal = Communications in Statistics. Simulation and Computation
\endbibitem

%b46 ###
\bibitem[\protect\citeauthoryear{Welling and Teh}{2011}]{welling2011bayesian}
%
\begin{binproceedings}[author]
\bauthor{\bsnm{Welling},~\bfnm{Max}\binits{M.}} \AND
\bauthor{\bsnm{Teh},~\bfnm{Yee~W.}\binits{Y.~W.}}
(\byear{2011}).
\btitle{Bayesian learning via stochastic gradient Langevin dynamics}.
In \bbooktitle{Proceedings of the 28th International Conference on
Machine Learning (ICML-11)}
\bpages{681--688}.
\end{binproceedings}
%
%\OrigBibText
%%
%\begin{binproceedings}[author]
%\bauthor{\bsnm{Welling},~\bfnm{Max}\binits{M.}} \AND
%\bauthor{\bsnm{Teh},~\bfnm{Yee~W}\binits{Y.~W.}}
%(\byear{2011}).
%\btitle{Bayesian learning via stochastic gradient Langevin dynamics}.
%In \bbooktitle{Proceedings of the 28th International Conference on Machine
%Learning (ICML-11)}
%\bpages{681--688}.
%\end{binproceedings}
%%
%\endOrigBibText
\bptok{imsref}%
\endbibitem

\end{thebibliography}
\end{document}